# Fair Railway Network Design

Zixu He[1*], Sirin Botan[1], Jérôme Lang[2*], Abdallah Saffidine[3],
Florian Sikora[4], Silas Workman[5]

[1] University of New South Wales
[2] CNRS, LAMSADE, Université Paris-Dauphine, PSL
[3] Potassco Solutions, Germany
[4] Université Paris-Dauphine, LAMSADE, PSL, CNRS
[5] Université Paris-Dauphine, PSL

tao.he@student.unsw.edu.au, {botan.sirin,abdallah.saffidine}@gmail.com,
{jerome.lang,florian.sikora}@dauphine.psl.eu, silas.workman@dauphine.eu

**Abstract**

When designing a public transportation network in a country, one may want to minimise the sum of travel duration of all inhabitants. This corresponds to a purely utilitarian view and does not involve any fairness consideration, as the resulting network will typically benefit the capital city and/or large central cities while leaving some peripheral cities behind. On the other hand, a more egalitarian view will allow some people to travel between peripheral cities without having to go through a central city. We define a model, propose algorithms for computing solution networks, and report on experiments based on real data.

## 1  Introduction

Long-distance inter-city travel is rapidly increasing (Schäfer et al. 2009), and is a major contributor to carbon emissions. In the past few years, there has been a significant shift in some parts of the world, with many people choosing train travel over flying for intra-continental trips. 2021 was the European Year of Rail, an initiative dedicated to promoting the use of train travel and how it can help Europe become carbon-neutral by 2050. Key to this goal is developing a railway network that prioritises minimising travel times for most users, subject to budget constraints.

Our aim in this paper is to contribute to this goal by leveraging notions of collective welfare from social choice, and combinatorial optimisation techniques. For this, we need to define a general model showing a trade-off between simplicity and realism. We propose a model for fair railway network design, with a parameterized collection of collective welfare functions, ranging from utilitarianism to egalitarianism, and to evaluate the impact of the fairness parameter on the shape of the resulting network. Assuming that each individual wants to travel between two cities, the cost incurred to an agent (an inhabitant) given a network is the shortest distance (assumed to be proportional to travel time) in the network between these two cities. These costs will be aggregated by a collective welfare function.

Using the notion of collective welfare enables us to be more precise in what we mean when we say "minimising travel times for most users". How do we aggregate the costs for individuals to define network optimality? A first idea would be minimising the *sum* of costs, which is known as an *utilitarian* approach. Utilitarianism is well-known to sacrifice some individuals (giving them a very high cost) against slightly decreasing the cost of a significantly larger number of individuals. The social choice literature argues for other ways of measuring social welfare, such as egalitarianism (minimising the maximum cost over all agents) or notions that fall between utilitarianism and egalitarianism. For this we consider a well-studied family of fairness measures, known as *sum of powers*: $p$ being a parameter in $\{1, 2, \ldots, +\infty\}$, if $c_i$ is the cost incurred by agent $i$, then the sum of powers aggregation of $(c_1, \ldots, c_n)$ is $\left(\sum_{i \in N} c_i^p\right)^{1/p}$ if $p$ is finite, and $\max_{i \in N} c_i^p$ if $p = +\infty$, where $N$ is the set of agents.[1]

It is reasonable to assume that utilitarianism will lead to networks that are more centralized, building more lines from/to cities that are bigger and more central, and neglecting peripheral, less populated cities (notably, by requiring people to travel between two peripheral cities to go through a central city), while more egalitarian criteria may lead to more balanced and less centralized networks, less detrimental to peripheral cities. For this, we posit a series of hypotheses, which we will then try to validate or invalidate on real-world data.

**Hypothesis 1** More egalitarian criteria result in more balanced networks. For this, we will consider well-known inequality indices, parameter $p$.

**Hypothesis 2** Everything else being equal, more central cities are better connected than peripheral cities, and the difference is less marked when $p$ increases.

---

[1] When working with positive utilities instead of costs, maximising the *Nash* social welfare (the product of utilities of all individuals) is known to be a sweet spot between utilitarianism and egalitarianism. However, it does not fit cost-minimising settings.

*Corresponding authors.

Obtaining a more balanced network is more fair: people living in the periphery should not feel they lose hours of travel in order to give a small gain to those living in the capital city. People living in the capital city are often used to train travel, and gaining a few minutes on each possible trip may not radically change their behaviour. On the other hand, people living in a peripheral mid-size city may not be as accustomed to train travel, and new lines to/from that city are more likely to have a significant impact. Therefore, we posit that being more egalitarian may lead to more people choosing the train, and therefore lower carbon emissions.

**Related work** Railway network design has been long studied in transportation economics and is known to be more complex than road network design because of its huge construction costs and the need to plan globally (Yi 2017). The design of a network typically relies on efficiency criteria (de Rus Mendoza 2012), such as market share, but not always (regional lines, for instance, often incur a deficit, are subsidised, and aim at reducing inequalities between territories). Transportation studies also aim at measuring the impact of factors that influence the choice of transportation (e.g., (Buehler 2011; Dütschke, an Anna Theis, and Hanss 2022)). Important determinants include price, travel duration and complexity, frequency, railway density, and past behaviour (Dütschke, an Anna Theis, and Hanss 2022).

For the case when $p = 1$, the problem has been studied from the algorithmic point of view under the name *Optimal Network Problem*: given a weighted graph and a budget, we look for a subgraph with a weight not exceeding the budget minimising the sum of weights of the shortest paths between all vertex pairs. This problem is shown to be NP-hard (Johnson, Lenstra, and Kan 1978) and different methods for solving it have been proposed (Dionne and Florian 1979; Boffey and Hinxman 1979) Approximating the problem better than $n^{1-\varepsilon}$ is also NP-hard, where $n$ is the number of cities (Wong 1980).

Social choice theory has a long history of evaluating solutions to collective decision problems by aggregating individual utilities by collective utility functions. Two well-known such functions are the utilitarian and egalitarian aggregation functions; some functions in between show a good balance between efficiency and fairness, such as Nash social welfare (for positive utilities), or sums of powers (Moulin 2004, Chap. 3).

An emerging field within in computational social choice is *participatory budgeting* (Aziz and Shah 2021; Rey and Maly 2023): given a set of candidate projects with individual costs, a maximal budget, and the preferences expressed by the citizens through votes, a subset of the project whose global costs does not exceed the budget should be selected in a fair way. Our problem seems to fall into this category (identifying projects with lines), and to some extent it does indeed, but with a major specificity: individual preferences over lines are not expressed directly and independently, but a set of lines is evaluated implicitly by the travel time it induces on the individual, making the evaluation of a set of lines by an agent highly non-additive. While there have been approaches to participatory budgeting with project interactions (Jain, Sornat, and Talmon 2020; Jain et al. 2021; Fairstein, Meir, and Gal 2021), the form of utility functions considered make them inapplicable to fair railway network design.

Finding a fair collective spanning tree connecting a set of nodes given an input graph has been investigated (Darmann, Klamler, and Pferschy 2009; Galand, Perny, and Spanjaard 2010; Darmann, Klamler, and Pferschy 2011; Escoffier, Gourvès, and Monnot 2013), with a major difference with our work: preferences bear over individual edges (and not over shortest paths).

**Outline of the paper** In Section 2 we present our model. Section 3 describes our algorithmic contributions: we show that the railway design problem is NP-complete for every value of $p$, present algorithmic techniques we use for computing optimal networks with respect to various aggregation functions, and discuss their performance. In Section 4 we discuss how the results obtained from running our algorithms on (simplified) real data allow us to verify how our assumptions are verified. We conclude in Section 5.

## 2 Model

At the basis of our model is an undirected, weighted graph, the *underlying network*. We have a set of *cities* $C$, with $|C| = n$; each city $C_i \in C$ has a population $p_i$. Further, $E \subseteq \{\{i, j\} \mid i, j \in C\}$ is a set of undirected edges between cities, with $|E| = m$, representing the possible direct lines that can be built.[2] Each edge $(i, j)$ in the $E$ has an associated weight $\ell_{i,j}$ representing the distance between the cities. We assume, for the sake of simplicity, that the cost of building a direct line between $i$ and $j$, and the travel duration along this line, are both proportional to $\ell_{i,j}$. For $(i, j) \in E$, when no direct line has been built between $i$ and $j$, it is also possible to travel between $i$ and $j$ at cost $K \cdot \ell_{i,j}$, where $K > 1$ is a constant. This is intended to express that in this case, agents may take a bus or private car along this line; this allows to output disconnected networks. Setting $K = \infty$ forces the network to be connected. For any pair of cities $i, j$, we define the *demand* function $\tau$ for trips between them as follows:

$$\tau_{i,j} : C \times C \to \mathbb{N},$$

where $\tau_{i,j}$ denotes the number of people who need to commute between $i$ and $j$. We require that demand is symmetric. In other words, we have that $\tau_{i,j} = \tau_{j,i}$.

Finally, we have a budget $B$, which is the amount of money available to construct any potential network.

---

[2]We allow for some pairs of cities to not be *directly* connected, because of geographical constraints. For instance, there cannot be a direct edge between Palermo and any continental Italian city: connections between Sicily and the continent have to be via Messina and Reggio Calabria.

A *railway network* $R \subseteq E$ is a set of edges chosen to be constructed. A railway network is *feasible* if the cost of building all edges in $R$ does not exceed the construction budget, i.e. $\sum_{e \in R} \ell_e \leq B$.

**Simplifying Assumptions**

In reality, railway construction as well as traveller preferences and costs depend on a huge number of variables. This means that we must make several simplifying assumptions in our model, in order to get off the starting line. Before proceeding further, we will discuss some of the main assumptions that we make in this paper.

Although real-world railway costs may vary depending on whether the line being built is a high-speed rail or not, here we consider only one type of line. We also require the cost of construction of a line between any two cities to be proportional to the distance between them. Again, this may not be the case in all real-world cases, for example, when geography prohibits the construction of a straight line.

We assume there is no preexisting railway network, which, while realistic in some cases, certainly does not model how all railway networks are built.[3] We also assume the budget is fixed and must be spent in one go.

**Population Needs and Costs** We assume that each citizen is interested only in commuting between two cities and that everyone commutes with the same frequency. While complexities such as population heterogeneity, the evolution of needs, and predicted behaviour of users (*e.g.*, how many will choose to travel by train between cities $A$ and $B$ if the travel time is 4 hours compared to a one-hour flight?) are a factor in real life, we do not consider them here.

We assume the time for an agent to travel along the edge $(i, j)$ is equal to $d_{i,j}$ (we ignore the proportionality factor) if there is a direct line between them, and $K \cdot d_{i,j}$ otherwise: we can think of this as the cost of taking a bus or driving along the edge $(i, j)$, rather than taking an indirect route via train. Finally, the cost of travelling from a city $i$ to a city $j$ is the weight of the shortest path between them. (This implicitly means that we ignore changes in lines, the cost in time and comfort associated with them, and the cost of switching from train to bus.)

**Travel Times**

The *travel time*, or *travel cost*, for an agent is the time needed to travel between their chosen pair of cities. Given a feasible network $R$, and $(i, j) \in E$, let $t_{i,j}$ be the travel time needed to traverse the edge $(i, j)$—when $(i, j) \in R$ then $t_{i,j} = d_{i,j}$, and otherwise $t_{i,j} = K \cdot d_{i,j}$.

Let $\hat{R}$ be the undirected weighted graph where each edge $(i, j)$ is weighted by $t_{i,j}$. For an agent wanting to travel between $i$ and $j$ in $C$, the travel cost $t_{i,j}^{\hat{R}}$ denotes the weight of the shortest path between $i$ and $j$ in $\hat{R}$.

**p-Egalitarian Social Welfare**

To measure social welfare, we define a parameterized social cost notion, the *p-Egalitarian Social Cost*, where the parameter $p$ indicates the degree of egalitarianism.

Given the demand function $\tau$, we define the *p-egalitarian social cost* of $R$ as follows

$$\text{SW}^p(R) = \left( \sum_{1 \leq i < j \leq m} \tau_{i,j} \left( t_{i,j}^R \right)^p \right)^{1/p}$$

When $p = 1$, $\text{SW}^1(R)$ will be equivalent to the utilitarian social cost (the sum of all individual costs); when $p = \infty$, we define $\text{SW}^\infty(R)$ as $\lim_{p \to \infty} \text{SW}^p(R) = \max_{1 \leq i < j \leq m} \left( t_{i,j}^R \right)$ which is the egalitarian social cost— i.e., the cost of the worst-off traveller.

**Instances**

An instance of the *railway design problem* is a tuple $(C, E, \ell, \tau, B, K)$.

Given a value of $p \in \mathbb{N}^+ \cup \{+\infty\}$, we define the *p-railway design problem* that takes as input an instance and a target social cost and returns whether there exists a feasible railway network with *p*-egalitarian cost lower than the target. Note that with $p = 1$ and uniform demand, this is equivalent to the Optimal Network Problem (Johnson, Lenstra, and Kan 1978).

**Example 1.** *We have one large city $X$, and two smaller cities $Y$ and $Z$. 16 persons want to commute between $X$ and $Y$, 16 between $X$ and $Z$, and 5 between $Y$ and $Z$ (shown in the figure in red). The distances between cities are 2 between $X$ and either of $Y$ and $Z$, and 1 between $Y$ and $Z$ (shown in the figure in blue).*

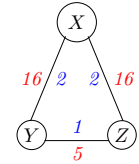

*A budget $B = 4$ allows us to build two edges so there are three maximal feasible networks $R_1$, $R_2$, $R_3$, in which we show $t_{i,j}$ in blue when it is along a train line and in orange when not (bus travel).*

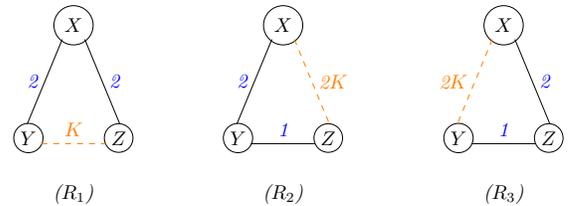

*We first fix $K = 5$. The travel times $t_{ij}^R$ given the three maximal feasible networks are shown in the table below, for $p = 1$, $p = 3$, and $p = \infty$.*

---

[3]For an example where this assumption matches reality, we can look to the construction of the Chinese high-speed network since the mid-2000s (Yi 2017, pp. 2–5).

|       | *Travel times* |         |         | *Social Cost* |           |              |
| :---: | :---: | :---: | :---: | :---: | :---: | :---: |
| $R$   | $t_{X,Y}^R$ | $t_{X,Z}^R$ | $t_{Y,Z}^R$ | $p=1$ | $p=3$ | $p=\infty$ |
| $R_1$ | 2 | 2 | 4 | **84** | $576^{1/3}$ | 4 |
| $R_2$ | 2 | 3 | 1 | 85 | $\mathbf{565^{1/3}}$ | **3** |
| $R_3$ | 3 | 2 | 1 | 85 | $\mathbf{565^{1/3}}$ | **3** |

*Note how the network that prioritises the citizens of the larger city $X$, is optimal for $p = 1$ while the other two are optimal for $p \geq 3$ and can be seen as more fair.*

*On the other hand, if we take $K = 3$ then for $R_1$ we have $t_{Y,Z} = 3$ (commuters between $Y$ and $Z$ will rather take the bus) and $R_1$ is optimal both for $p = 1$ and $p = 3$, and all three networks are optimal for $p = \infty$.*

## 3 Identifying Good Networks

Now that optimal networks for $p$ have been defined, how can they be computed?

**Computational Hardness**

Recall that the Optimal Network Problem mentioned in the introduction is equivalent to the $p$-railway design problem when $p = 1$ and demand is uniform. The Optimal Network Problem is known to be (strongly) NP-complete (Johnson, Lenstra, and Kan 1978). We will show that this remains true for all values of $p$.

**Proposition 1.** *For any fixed $p \in \mathbb{N}^+ \cup \{+\infty\}$, the $p$-railway design problem is strongly NP-complete.*

*Proof sketch.* Membership in NP is straightforward. For hardness, we reduce from SAT and provide a detailed proof in Appendix A. The main idea is to create *clause* cities, *literal* cities, and a distinguished *capital* city. There are edges between each clause city and the cities corresponding to literals in it, and between each literal city and the capital. The distances and the construction budget are set so that all clauses can be directly connected to their literals, and exactly half of the literals can be directly connected to the capital. The demand and the target social cost are chosen such that a network is a solution if and only if every clause can reach the capital city in exactly two steps. □

Not only is the exact optimization problem hard, but for $p = 1$, it is also NP-hard to approximate within $n^{1-\varepsilon}$ (Wong 1980) [4]. These are bad news; however, we still need to solve the problem in practice to analyze some networks. To this end, we will follow two parallel approaches:

- For large networks, since it is impractical to develop an exact algorithm, we use a heuristic algorithm based on local search, with a sophisticated initialization step that includes edges with high marginal contributions to decreasing social cost.

---

[4]With $p = 1$, when restricting solutions to spanning trees, the problem is still (surprisingly enough) NP-hard, but admits efficient approximation algorithms. (Wu et al. 1999)

- For small networks (up to 13 cities), we use an exact optimization algorithm based on tree search with constraints encoded in Answer Set Programming.

We will use the exact algorithm to validate the heuristic algorithm by comparing the solutions it provides for up to 13 cities with the optimal solutions. We begin by describing the local search algorithm.

**Local Search Algorithm**

The local search algorithm is composed of three stages:

1. *preprocessing*: edges that are not promising are filtered out, resulting in a subset of edges $E^*$.
2. *generation of the initial network*: $E^*$ is further reduced by filtering out more edges until a feasible network $R_0$ is obtained.
3. *local search improvements*: starting from $R_0$, a series of local improvements is made by replacing a small number of edges in the current network $R$ with another small number of edges from $E^*$. The process stops when a local optimum $R^*$ is reached.

Stages 1 and 2 are based on a heuristic function called the *marginal contribution of edges*. The $p$-marginal contribution of an edge between two cities, given a network, is the $p$-social cost increase resulting from excluding the edge from the network: for each edge $e$, we define

$$\text{MC}(e, R, p) = \frac{\text{SW}^p(R - \{e\}) - \text{SW}^p(R)}{t_e^R},$$

where $R - \{e\}$ is the network obtained by removing edge $e$ from network $R$.

When $p = 1$, marginal contribution is similar to the $\gamma$-*values* defined by Boffey and Hinxman (1979) (for a different purpose), with a key difference: $\gamma$-values are computed only by considering agents who travel along an edge $e$ and ignoring other agents (who can still benefit from $e$). In other words, $\gamma$-values (resp. marginal contributions) have a local (resp. global) flavour.

Intuitively, an edge has a large marginal contribution when the travel demand between the cities it links is high, and there exists no short alternate route between them. Edges with a small marginal contribution are likely to be less important and are therefore considered first for deletion.

**Example 2.** *Consider the instance below where distances are displayed on the graph and the demand between cities is equal to 1 for every pair of cities. We take $K = 3$.*

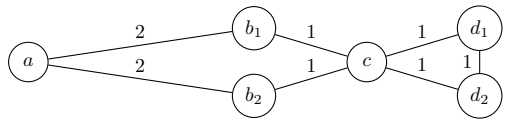

*The MC values are given in the table below.*

| $(u, v)$ | $SW^p(R - \{uv\})$ | $d^R(u, v)$ | $MC(uv, R, p)$ |
| :---: | :---: | :---: | :---: |
| $(b_i, c)$ | 43 | 1 | 12 |
| $(a, b_i)$ | 33 | 2 | 1 |
| $(c, d_i)$ | 35 | 1 | 4 |
| $(d_1, d_2)$ | 32 | 1 | 1 |

*Let us give the details for $(b_1, c)$. The social cost for $R$ is 31 (1 for each $b_i c$ and $c d_i$, 2 for each $a b_i$, $b_i d_j$, as well as $b_1 b_2$ and $d_1 d_2$, 3 for ac, and 4 for each $a d_i$). If we remove $(b_1, c)$ from $R$ the the impacted costs are for $b_1 c$ (5 instead of 1), $b_1 d_1$ and $b_1 d_2$ (6 instead of 2)—a cost increase of 12, to be divided by $d_{b_1, c} = 1$.*

The marginal contribution will be used for three different purposes: (1) pre-processing by filtering edges before launching an algorithm on large instances, (2) generating a feasible network, and (3) serving as a branching heuristic for the exact optimization algorithm.

**Preprocessing** The first stage of the local search method involves filtering out edges until the set of available edges $E^*$ reaches a fixed number $\sigma$.

To compute $E^*$, we use the following procedure, which iteratively identifies the edge with the lowest marginal contribution and removes it from $E$ until $\sigma$ edges remain. $E^*$ is then defined as the resulting set $E$.

---

1: **for** $i = \sigma + 1$ **to** $m$ **do**
2: $\quad e^* \leftarrow \arg\min_{e \in E} \text{MC}(e, E, p)$
3: $\quad E \leftarrow E - \{e^*\}$
4: **end for**
5: $E^* \leftarrow E$

---

**Greedy Network Generation** In a greedy network generation process, we start with the set of edges $E^*$ and iteratively eliminate edges—based on their dynamic marginal contribution—until we reach a feasible network that does not exceed the budget, which we denote by $R_0$.

**Local Search Improvements** The local search algorithm starts from $R_0$ and finds a series of local improvements that reduce $p$-social cost until a local optimum is reached.

Let us fix a value of $p$, and the set of allowed edges $E^*$. Given two integers $m_{\text{add}}$ and $m_{\text{del}}$, and a feasible network $R$, an $(m_{\text{add}}, m_{\text{del}})$-*local improvement* is a feasible network $R'$ such that (i) $\text{SW}^p(R') < \text{SW}^p(R)$, (ii) $R'$ is feasible, (iii) $R'$ is obtained by removing at most $m_{\text{del}}$ edges from $R$ and adding at most $m_{\text{add}}$ edges from $E^* - R$.

The local search algorithm $\text{LS}(R, m_{\text{add}}, m_{\text{del}})$ outputs the local optimum reached after performing a series of $(m_{\text{add}}, m_{\text{del}})$-local improvements.

**Exact Algorithm**
This algorithm takes as input an instance of the railway design problem and outputs an optimal network for a fixed value of $p$. Before the search, a preprocessing step ranks all edges (pairs of cities) according to their marginal contribution. The algorithm maintains a current optimal network $R^*$ (the minimal cost feasible network found so far) together with its cost $c^*$, and repeatedly uses the following improvement subroutine: given an instance and a cost $c^*$, find a feasible solution with a cost strictly smaller than $c^*$, until no further improvements are possible. This subroutine is based on a brute-force search with pruning, leveraging constraint propagation using answer set programming (ASP) (Lifschitz 2008). The feasible solutions with a cost less than $c^*$ coincide with the models of the ASP program.

The ASP solver branches on logical variables corresponding to edges, using a domain ordering heuristic (Gebser et al. 2013): it branches on edge inclusion in the network, starting from the edge with the highest marginal contribution down to the edge with the least marginal contribution, as identified in preprocessing. The program comprises (1) constraints to eliminate over-budget models (those including too many edges) and (2) constraints to eliminate over-cost networks. Constraints of type (1) are expressed with aggregators, while constraints of type (2) are enforced on the fly with a theory propagator (Gebser et al. 2016). Whenever the solver identifies a solution candidate $R \subseteq E$ (corresponding to a leaf of the search tree), the propagator computes its social cost. If it is less than $c^*$, then $R$ is indeed a solution, and the subroutine concludes. Otherwise the propagator adds $\bigvee_{e \notin R} e$ as a new clause to the logic program. This new constraint eliminates $R$ and any subset of $R$. The cost of any network thus eliminated would have been no smaller than $c^*$ because the social cost is non-decreasing with respect to edge deletion. [5]

The running times for the exact algorithm are shown in Figure 1a. The program allows us to compute the exact optimal networks for all considered values of $p$ and up to 13 cities.

**Experimental Validation**
For the implementation of the local search algorithm, we first run the algorithm with $m_{\text{add}} = 1$ and $m_{\text{del}} = 2$ to obtain a medium-quality network $R'$. Then, we run the local search algorithm starting from $R'$ with $m_{\text{add}} = 2$ and $m_{\text{del}} = 2$ to obtain the final network $R''$. In the first run, the network converges to a local optimum with a faster running time, while in the second run, the network converges to a better local optimum with a slower running time.

We demonstrate that the local search algorithm is close to the optimum (as given by the exact algorithm) for a small number of cities and, as expected, is more scalable. To illustrate this, we compare the running time and solution quality of the local search and exact algorithms. Both algorithms were run on the same maps and budgets. Using France as an example, Figure 1a shows the average running time for both algorithms and the average ratio of the social cost[6] between the local

---

[5] The code is available at https://github.com/HtBest/Fair-Railway-Network-Design-code.

[6] For example, if the local search outputs the optimal network, then the ratio is 1; if the social cost of the output

search output and the exact algorithm output, with up to 13 cities on 60 sampled budgets.[7] The results show that the local search algorithm is more scalable and produces solutions of similar quality to the exact algorithm. It can be observed that the ratio increases with the number of cities. However, even with 13 cities, on average, local search only adds 1‰ to the social cost. Results for other countries are similar.

Since the social cost of networks identified by the local search algorithm is very close to the optimal solution and its running time is much faster compared to the exact algorithm, we will use the local search algorithm for the remaining experiments in this paper.

## 4 Results and Discussion

We now assess the validity of our hypotheses in light of the networks obtained from realistic input data.

The experimental study will compare the obtained networks and their quality with respect to: (a) the cost distribution (commuting time) across all individuals; (b) the distribution across all cities $c$ of (b1) the number of lines incident to $c$, and (b2) the average cost for all agents living in $c$.

### Input Data

We considered nine countries spread across the globe with significantly different sizes, populations, and distributions of population: Brazil, Canada, France, Germany, Italy, Russia, Spain, the UK[8], and the USA. For each country, we considered the largest 20 urban areas. We assume that the number of people who want to commute between any two cities is proportional to the product of their population divided by the distance between them; this is a known reasonable demand model (starting with (Zipf 1946)). We fix the constant $K = 3$. We then compute the networks using the algorithm discussed in Section 3 for different values of $p$. We choose five values of $p$: 1, 2, 4, 6, and 10.[9]

The analysis in this section is based on the networks obtained using the local search algorithm discussed in Section 3. For the sake of variety, we will use figures for different countries to illustrate our findings; all figures for all countries are in the appendix.

### How does $p$ influence fairness?

We first assess the level of fairness in the output networks by computing the Gini index of the cost (travel time) distribution for all inhabitants. The Gini index is a statistical tool that measures inequality (its formal definition is provided in Appendix B); a higher index indicates greater inequality among inhabitants and a less balanced network.

Figure 1b shows the evolution with the available budget of the Gini index for the UK networks with 20 cities, for $p = 1, 2, 4, 6, 10$, starting from a budget just enough to build one edge to the budget that can build all edges.

For all countries, the Gini index not only tends to decrease asymptotically with the budget for all values of $p$, but the differences between different values of $p$ also tend to vanish. This is because large budgets lead to near-optimal networks, which are already quite egalitarian even for $p = 1$. When the budget is sufficient to build all edges, the Gini index for different values of $p$ converges to the same value, which we refer to as the baseline. This baseline is independent of both $p$ and $K$, and is thus specific to each country.

We observe that larger values of $p$ lead to less inequality (lower Gini index), which was expected. More surprising, however, is the observation (consistent across all countries) that the curve for $p = 1, 2$ almost always stays above the baseline, while the curve for $p = 4, 6, 10$ almost always remains below it. We currently do not have a clear explanation for this phenomenon.

An exciting output consists of the different networks obtained for some countries when the budget or $p$ varies. Space prevents us from giving a variety of examples so we pick one country (France) and one budget (2890) and show the networks obtained for $p = 1, 4$, and $10$.

We give a few comments on the three networks depicted in Figure 2. We observe that for $p = 1$, Paris (by far the largest city) is a hub, and when $p$ increases, the number of direct lines to Paris decreases and the new hub becomes Clermont-Ferrand, making the network more fair by allowing better connections between Northwest and Southeast and between Northeast and Southwest. Building expensive lines allowing this (such as Clermont-Nantes or Clermont-Nancy) is made at the expense of connectedness: cities that are close to a well-connected city such as Rouen (close to Paris), Saint-Etienne or Grenoble (close to Lyon) are now disconnected to the rest of the network and have to commute via Paris or Lyon. This behaviour is somewhat specific to the choices of a small budget $B$ (which implies that a fully connected network tends to be a spanning tree, as is the case for $p = 1$) and a moderate value of $K$ ($= 3$), allowing people commuting to/from disconnected cities near well-served cities to be still well-satisfied. With a smaller value of $K$ and a small budget, we would observe more disconnections (even for $p = 1$) and for large values of $K$, the network tends to be connected for all values of $p$. Many more examples are in the Appendix D.

### The Influence of Centrality

We now turn to Hypothesis 2. The *average cost of a city i for network R* is the average travel time w.r.t. $R$

---

network is 1.01 times the optimal cost, the ratio is 1.01.

[7]The programs were run on an Apple Macbook Pro with an M3Max CPU and 48GB RAM, using Python3.11.

[8]Of course we do not include Northern Ireland, which is disconnected from the rest of the country. Still, we will say "the UK" and not "Great Britain".

[9]We do not consider higher values, as $p = 10$ has already a very egalitarian flavor. We computed the optimal networks for $p = \infty$, but we don't report on them here (we do so in the appendix), as they don't pay attention at all to city populations, they go too far in egalitarianism.

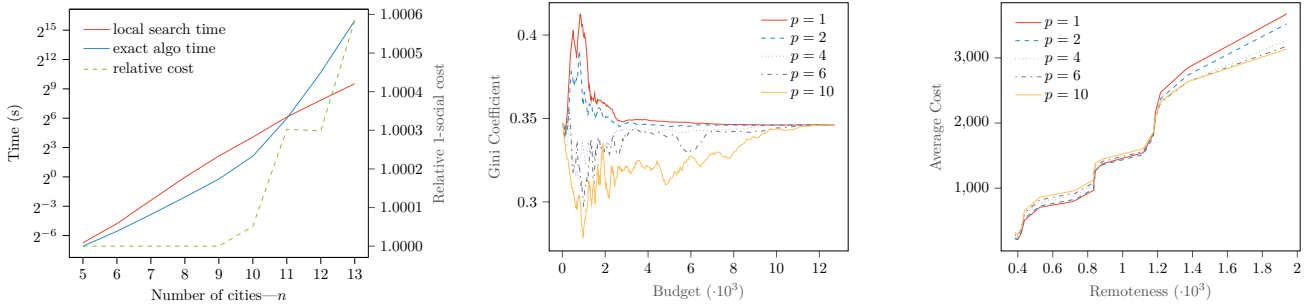

(a) France, $p=1$, running time (log scale) for both algorithms and the average ratio of social cost of local search to the optimum.

(b) The UK, the change of Gini index with budget under different values of $p$.

(c) Brazil, the impact of the remoteness of cities on the average cost for each city with different $p$.

Figure 1: Statistical results

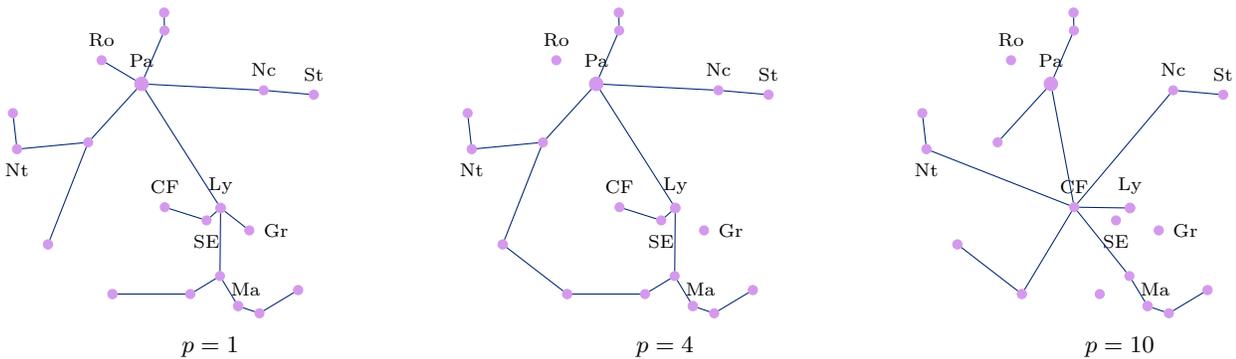

Figure 2: Solution networks for France with budget 2890, $K=3$

of people who commute to/from $i$:

$$\mathrm{AC}(i, R) = \frac{\sum_{j \neq i} \tau_{i,j} d^R_{i,j}}{\sum_{j \neq i} \tau_{i,j}}$$

For a given underlying network, for each value of $p$, we sample 300 possible budgets. For each sampled budget $B$ we determine the optimal network $R_p(B)$ and compute $\mathrm{AC}(i, R_p(B))$; The varying-budget average cost $\mathrm{VAC}(i)$ of $i$ is the average of the values $\mathrm{AC}(i, R_p(B))$ obtained for the 300 sample budgets.

Now, we define the *remoteness* of a city as follows (the more remote a city, the less central it is):

$$\mathrm{Re}(c) = \frac{\sum_{i \in C \setminus \{c\}} \ell_{c,i}}{|C|-1}.$$

Figure 1c shows how the average commuting cost $\mathrm{VAC}(i)$ varies with remoteness for different values of $p$ in Brazil. We observe that (1) central cities tend to have lower average commuting costs for all values of $p$ (which was of course expected) and, more interestingly, that (2) this advantage decreases with $p$ (lower $p$ value favours the central cities, while higher $p$ value favours the less central cities), which is consistent with Hypothesis 2. For Brazil, for cities that have a remoteness value less (respectively, more) than 1200, lower (respectively, higher) $p$ values tend to give a lower average commuting cost. The threshold remoteness values are different for different countries, but the trend is consistent.

Of course, centrality is not the only factor that explains that a city is better off with lower or larger values of $p$. Another factor is, obviously (because of the definition of $p$-social costs), city populations (with $p = \infty$ as an extreme, where populations don't count at all).

## 5 Conclusion and Future Work

We have paved the way towards the use of social choice-theoretic fairness notions for the design of transportation networks. Our main contribution is a model, algorithms for computing optimal or approximately optimal networks, and the analysis of the experimental results we obtained using the simplified map of nine countries.

What remains to be done includes (1) defining a more refined model with several types of lines and pre-existing networks; (2) obtaining results from realistic simulated maps; (3) doing a reverse study, where we start from existing networks and compare their $p$-social cost according to various of $p$; this will help us locate existing networks on the utilitarian-egalitarian scale.

# A  NP-hardness proof

**Proposition 1.** *For any fixed $p \in \mathbb{N}^+ \cup \{+\infty\}$, the p-railway design problem is strongly NP-complete.*

*Proof.* We give a polynomial reduction from the NP-hard problem 3-SAT. An instance $\mathcal{I} = (X, D)$ of 3-SAT is a set of Boolean variables $X$ such that $|X| = n$, and a collection of clauses $D$ such that $|D| = m$, where a clause is a disjunction of exactly 3 literals taken from $X$ (a variable or a negation of a variable). The 3-SAT problem asks whether there exists an assignment of each variable of $X$ such that each clause is satisfied.

Given an instance $\mathcal{I}$ of 3-SAT, we build an instance of the decision version of the $p$-railway design problem as $\mathcal{J} = (C, E, \ell, \tau, B, K, \text{SW})$ as follows (see Figure 3). We define $k = mn + 2n + 1$.

$$C = \{v_x, p_x, n_x \mid x \in X\} \cup \{c_d \mid d \in D\} \cup \{a\}$$
$$E = \{(a, p_x), (a, n_x), (v_x, p_x), (v_x, n_x) \mid x \in X\}$$
$$\cup \{(c_d, p_x) \mid d \in D, x \in d\}$$
$$\cup \{(c_d, n_x) \mid d \in D, \neg x \in d\}$$
$$\ell_e = \begin{cases} k & \text{for } e \in \{(a, p_x), (a, n_x) \mid x \in X\} \\ 1 & \text{for } e \in E \setminus \{(a, p_x), (a, n_x) \mid x \in X\} \end{cases}$$
$$\tau_e = \begin{cases} 1 & \text{if } e \in \{(a, v_x), (a, c_d) \mid x \in X, d \in D\} \\ 0 & \text{otherwise} \end{cases}$$
$$B = 2n + 3m + nk$$
$$K = +\infty$$
$$\text{SW} = (m+n)^{1/p}(k+1)$$

Informally, we have a classical incidence graph of the 3-SAT instance. We add moreover a vertex for each variable, connected to both its literals. We also add a vertex $a$, connected to all literal vertices, but with a high cost. The demands are only from vertex $a$ to all clause vertices and from $a$ to all variable vertices. The budget is set-up such that all edges of weight 1 can be built $(2n+3m)$ in addition to one edge from $a$ to exactly one literal for each variable $(kn)$. Note that when $p$ is set to $+\infty$, SW equals $k + 1$.

We claim that $\mathcal{I}$ is satisfiable if and only if there is a $R \subseteq E$ respecting the budget $B$ and having a social cost of at most SW.

Assume that $\mathcal{I} = (X, D)$ is satisfiable and let $f$ be an assignment to the variables of $X$. We build $R \subseteq E$ as follows : add all edges with $l_e = 1$, and add the edge $(a, p_x)$ if $f(x)$ is true, otherwise add the edge $(a, n_x)$. One can easily check that it costs exactly $2n + 3m + kn = B$. The social cost for each demand $(a, v_x), x \in X$ is of $(k+1)^p$ (with path going through either $p_x$ or $n_x$ according to the assignment of $x$), and for each $(a, c_d), d \in D$ is of $(k+1)^p$, with a path going through one literal satisfying $d$, which exists since $\mathcal{I}$ is satisfiable. Therefore, the total social cost is $(m(k+1)^p + n(k+1)^p)^{1/p} = \text{SW}$

For the reverse direction, assume that $R \subseteq E$ is a solution to $\mathcal{J}$. First, observe that the length of a path between all pairs of cities with positive demand is at least $k + 1$, and that it is also the maximum possible length in order to respect SW – therefore, the length of this path must be exactly $k+1$ in any positive solution. From the previous observation, we can deduce that for every demand $(a, v_x), x \in X$, we must have exactly one of the edges $(a, p_x), (a, n_x), x \in X$. Indeed, having none would gives a path of length strictly greater than $k+1$ and thus exceed the social cost SW and having both would exceed the budget. Therefore, we can assign $x$ to True if $(a, p_x) \in R$, and $x$ to False otherwise.

For the demands $(a, c_d), d \in D$, the length of the corresponding path must also be $k+1$, which is possible if and only if there is an edge $(a, l_x)$ in the solution, for some literal $l_x \in d$. By our previous assignment of the variables, such path exists only if the assignment satisfies the clause.

Since all weights are at most $k$, which is polynomially bounded by the size of the SAT formula, it concludes the claimed result of strong NP-hardness.

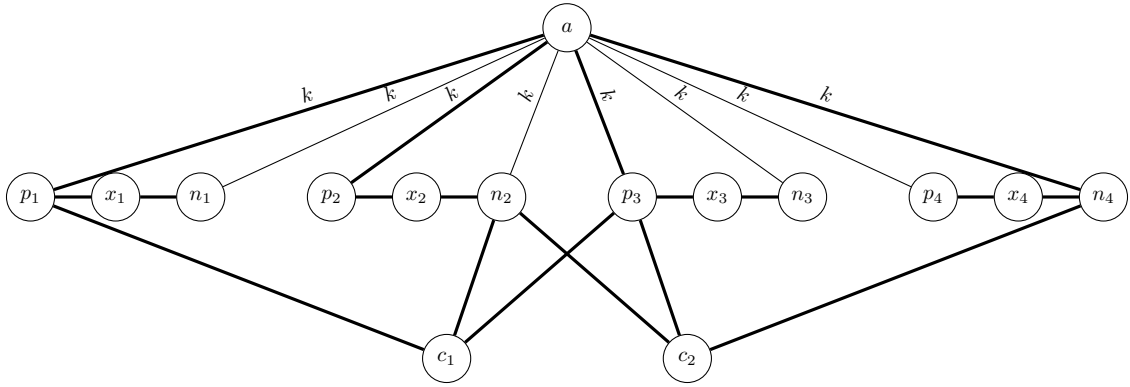

Figure 3: Example of the reduction from a toy instance $(x_1 \vee \neg x_2 \vee x_3) \wedge (\neg x_2 \vee x_3 \vee \neg x_4)$. Edges without label are of cost 1. Bold edges correspond to the built solution considering $x_1 = x_2 = x_3 = True$ and $x_4 = False$.

## B  Figures: fairness

**Gini index**

The Gini index (Gini 1936) is a statistical tool that measures economic inequality between individuals in a group (such a country or a region). In economics, the Gini index is often applied to wealth or revenue. Here we apply it to travel distances. The Gini index of a network $R$ under instance $\mathcal{I}$ defined as follow:

$$G = \frac{\sum_{1 \leq i < j \leq n} \sum_{1 \leq k < l \leq n} \left| t^R_{i,j} - t^R_{k,l} \right| \tau_{i,j} \tau_{k,l}}{2 \sum_{1 \leq i < j \leq n} \sum_{1 \leq k < l \leq n} t^R_{k,l} \tau_{i,j} \tau_{k,l}}$$

The values of the Gini index range from 0 (perfect equality) to 1 (extreme inequality).

We give below the figures for all 9 countries considered, that show the evolution if the Gini index or the output network with a budget varying from small to large, and for different values of the fairness parameter $p$: Figures 4a for Brazil, 4b for Canada, 4c for France, 4d for Germany 4e for Italy 4f for Russia 4g for Spain 4h for the UK, and 4i for the USA.

**Worst-best ratio**

The Gini index is not the only way to measure how unequal the network is. In particular, if the network is fair to almost all cities, except for one or two cities that are extremely disadvantaged, then the Gini index will be reasonably low, although the network is arguably not fair. Another way of measuring inequality consists simply of computing the ratio between the average travel costs of the worst and best cities. This is depicted on Figures 5a, 5b, 5c, 5d, 5e, 5f, 5g, 5h and 5i.

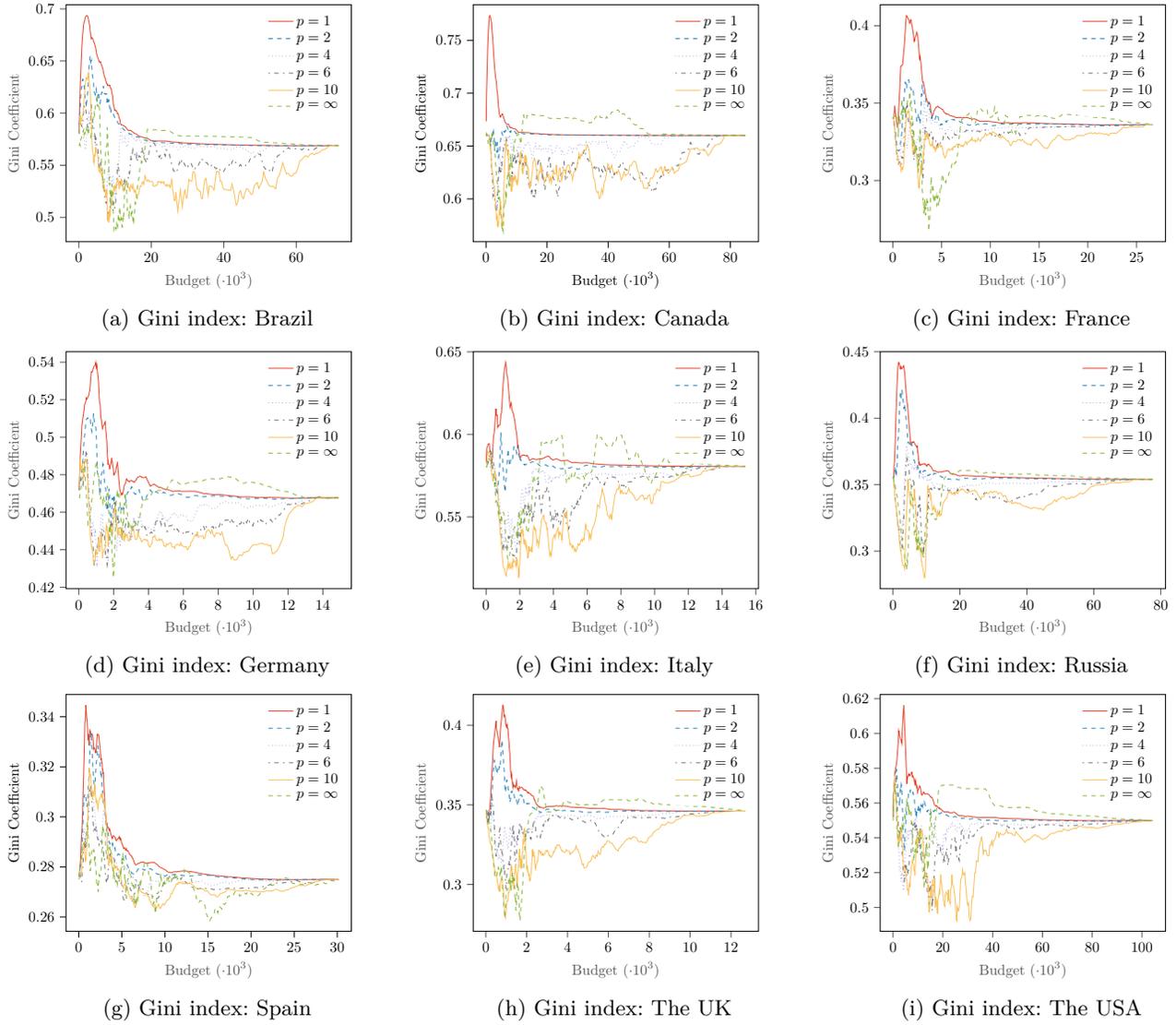

Figure 4: Evolution of the Gini index with budgets for different values of $p$

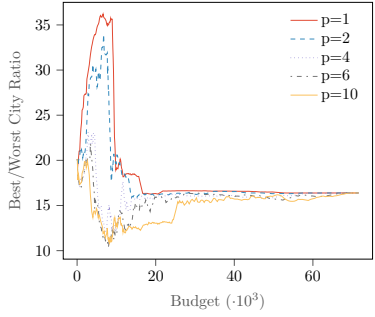

(a) Average travel cost ratio: Brazil

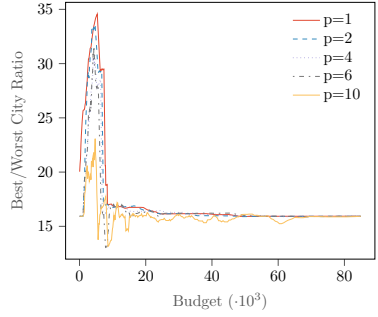

(b) Average travel cost ratio: Canada

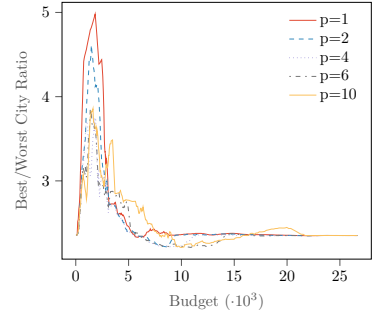

(c) Average travel cost ratio: France

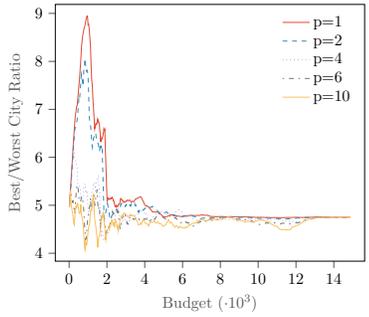

(d) Average travel cost ratio: Germany

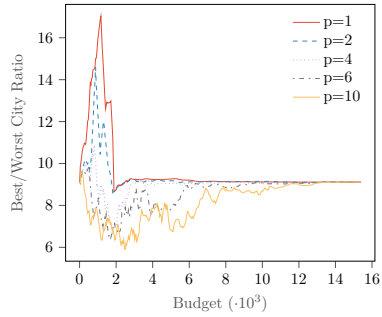

(e) Average travel cost ratio: Italy

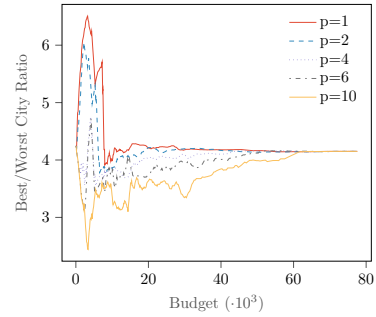

(f) Average travel cost ratio: Russia

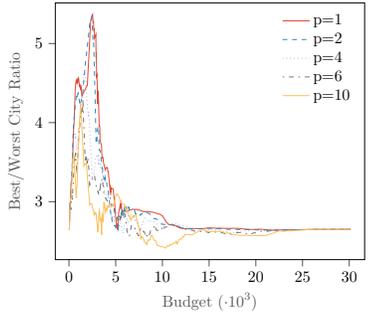

(g) Average travel cost ratio: Spain

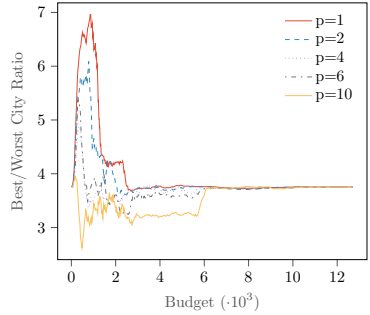

(h) Average travel cost ratio: The UK

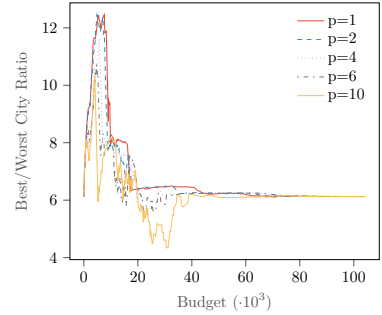

(i) Average travel cost ratio: The USA

Figure 5: Evolution of the ratio of average travel cost between the worst and best cities with budgets for different values of $p$

## C  Figures: influence of remoteness

Now we present figures that show the influence of centrality (or remoteness) of cities on the average travel cost of the individuals who travel from/to that city, depending on the fairness parameter $p$.

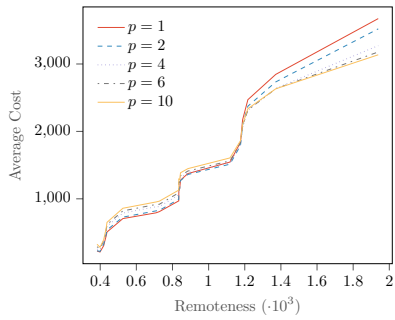
(a) Brazil

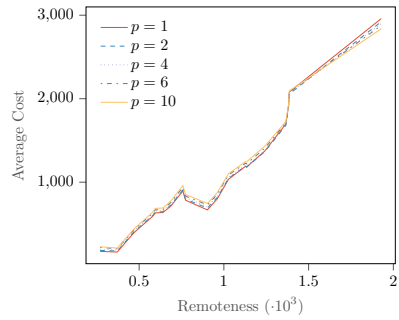
(b) Canada

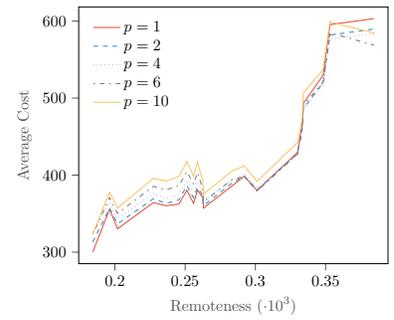
(c) France

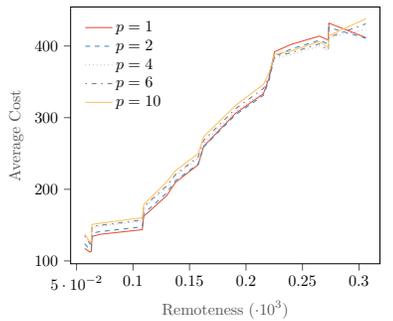
(d) Germany

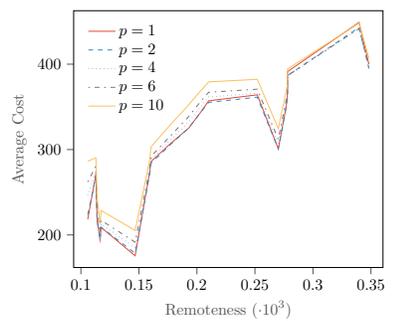
(e) Italy

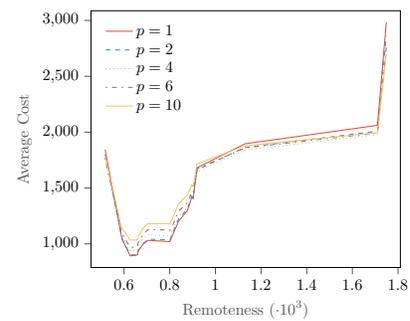
(f) Russia

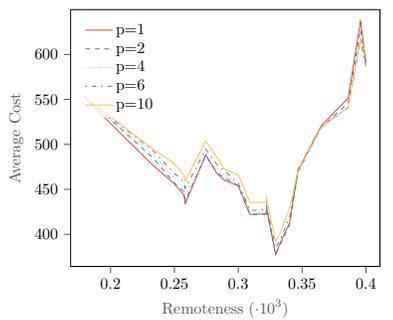
(g) Spain

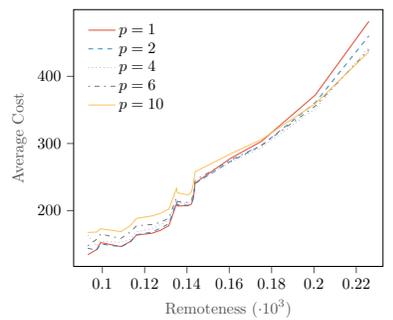
(h) The UK

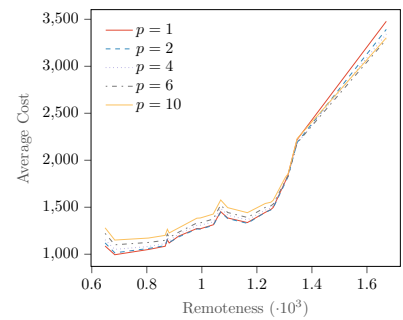
(i) The USA

Figure 6: The change of average cost with remoteness under different values of $p$

## D   Output networks

Now we present a selection of output networks, for each of the 9 countries considered, and for different values of $p$, $B$, and $K$.

Note that some of the maps (especially Canada) look very different from what they usually do. This is because we always normalize the position to a square; if we don't then we will get very unreadable flat maps where all edges are squeezed together.

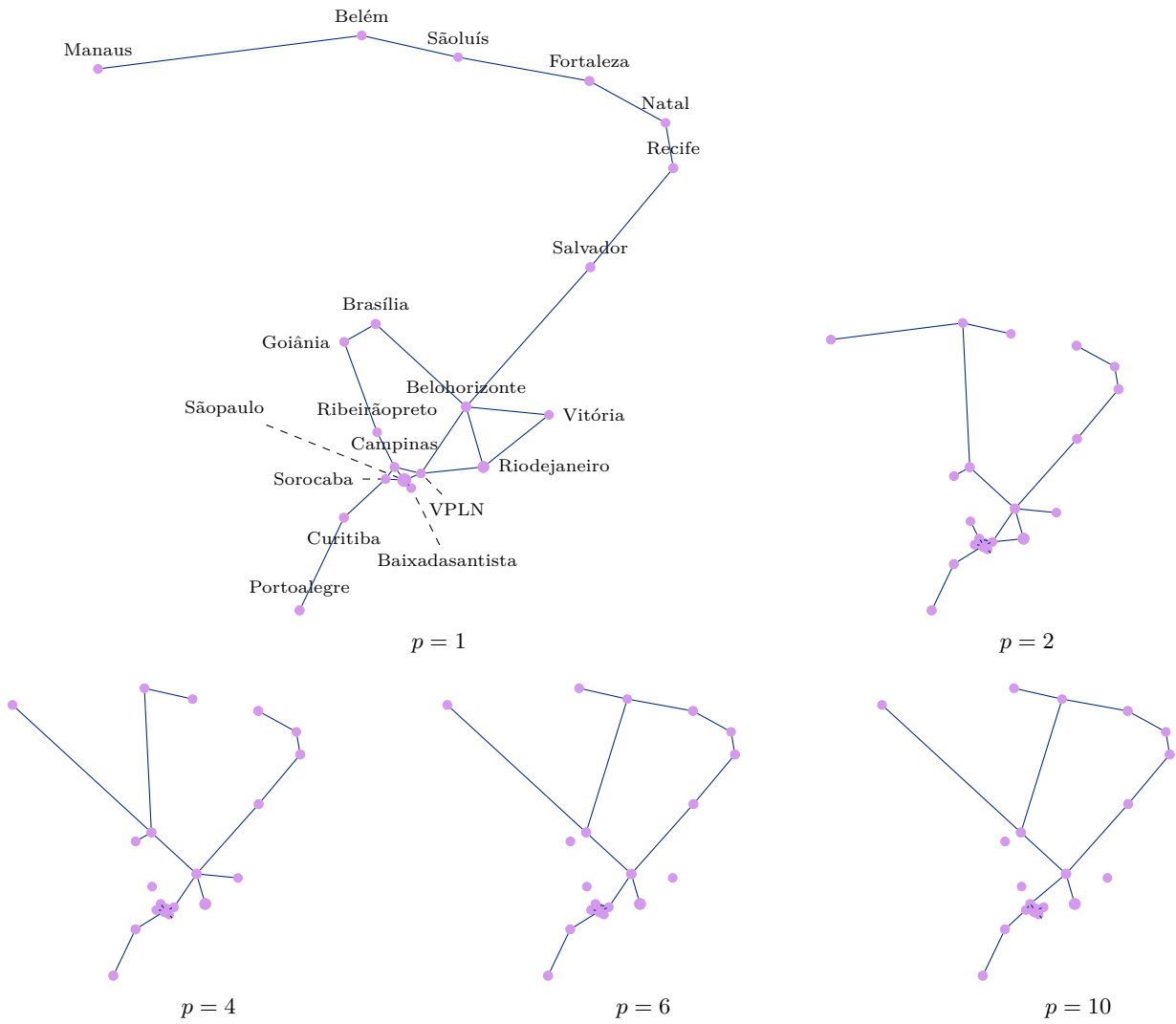

Figure 7: Solution networks for Brazil with budget 9510, $K = 3$

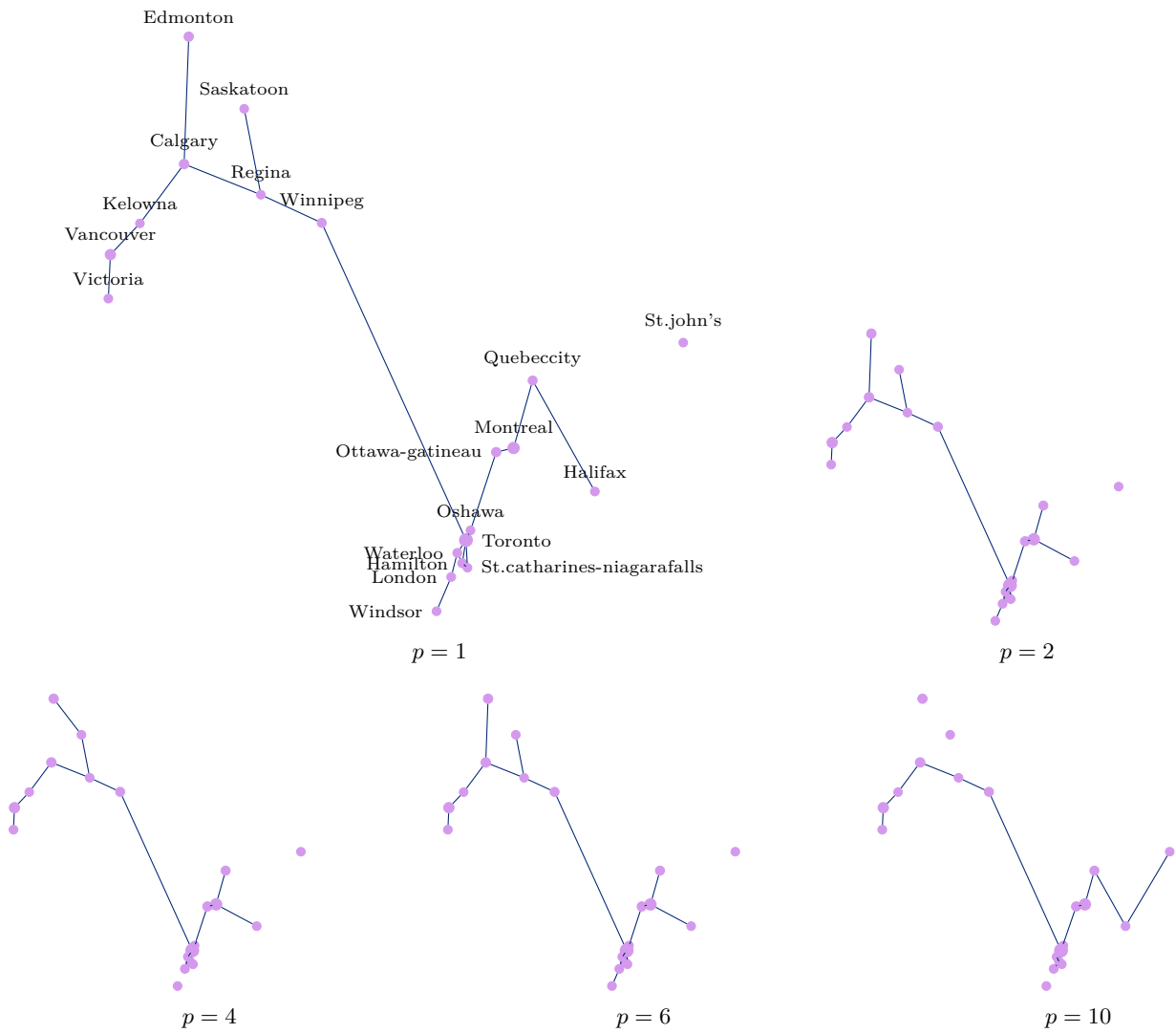

Figure 8: Solution networks for Canada with budget 6003, $K = 3$

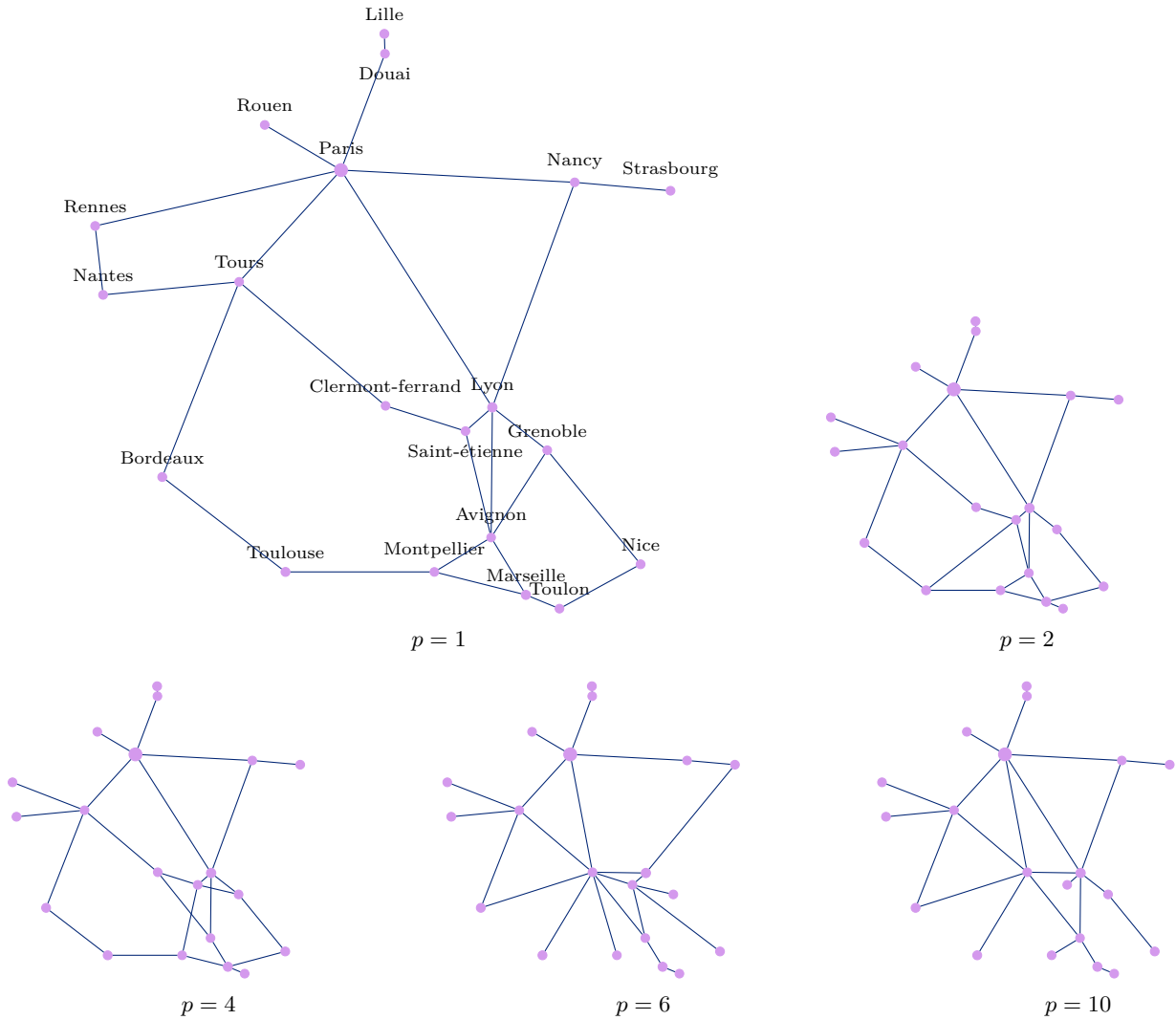

Figure 9: Solution networks for France with budget 4642, $K = 3$

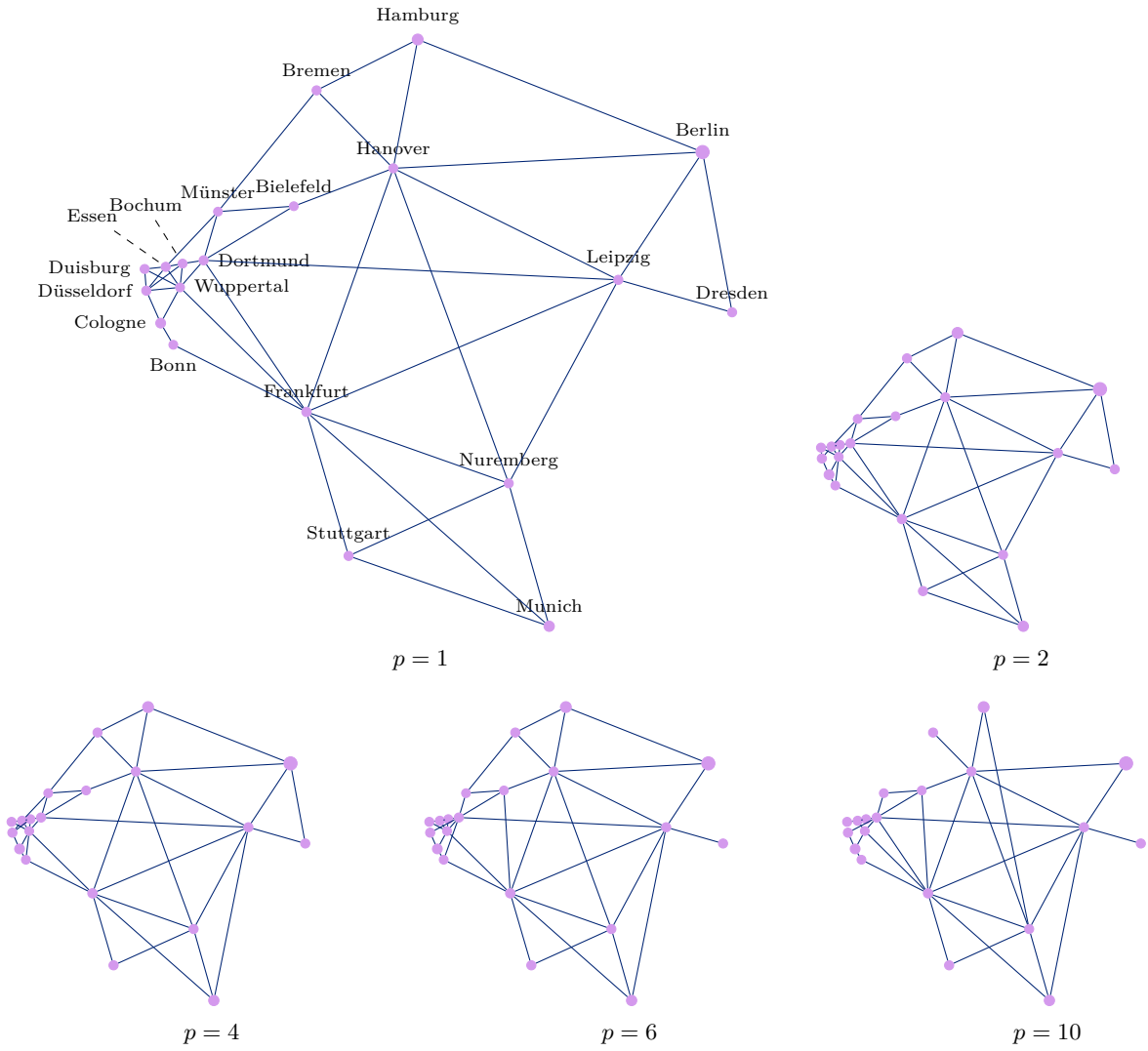

Figure 10: Solution networks for Germany with budget 5446, $K = 3$

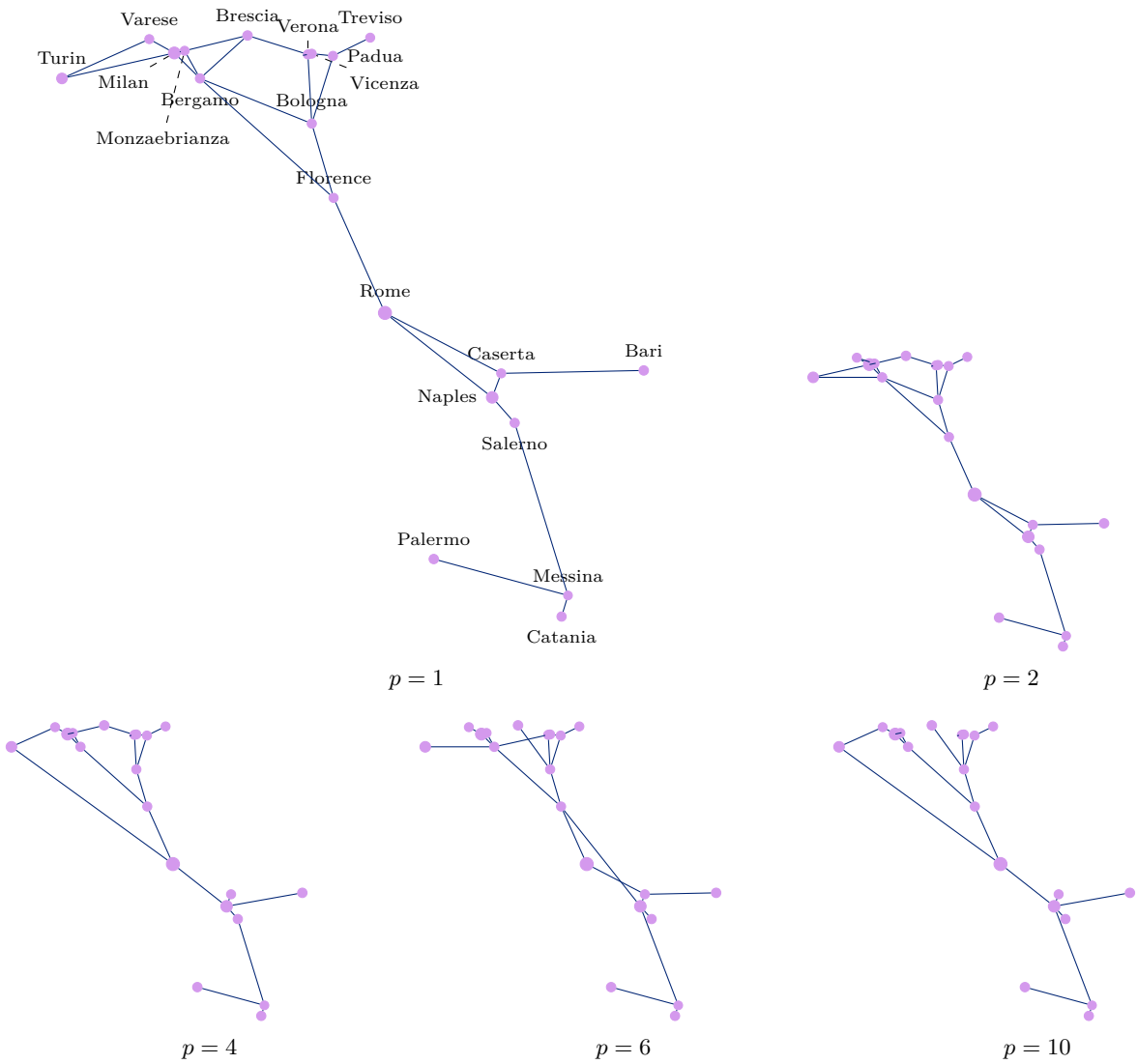

Figure 11: Solution networks for the Italy with budget 2813, $K = 3$

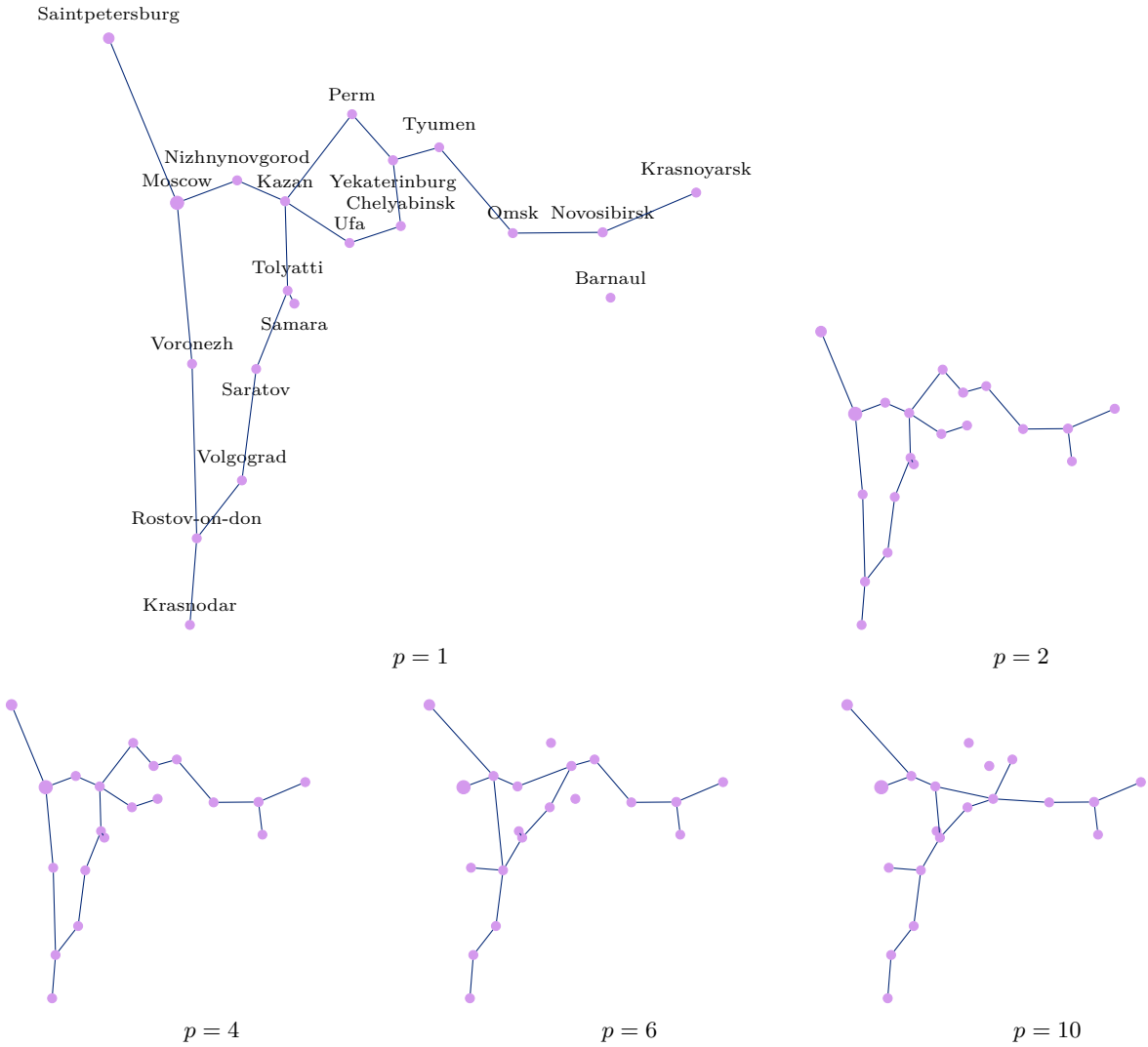

Figure 12: Solution networks for Russia with budget 7828, $K = 3$

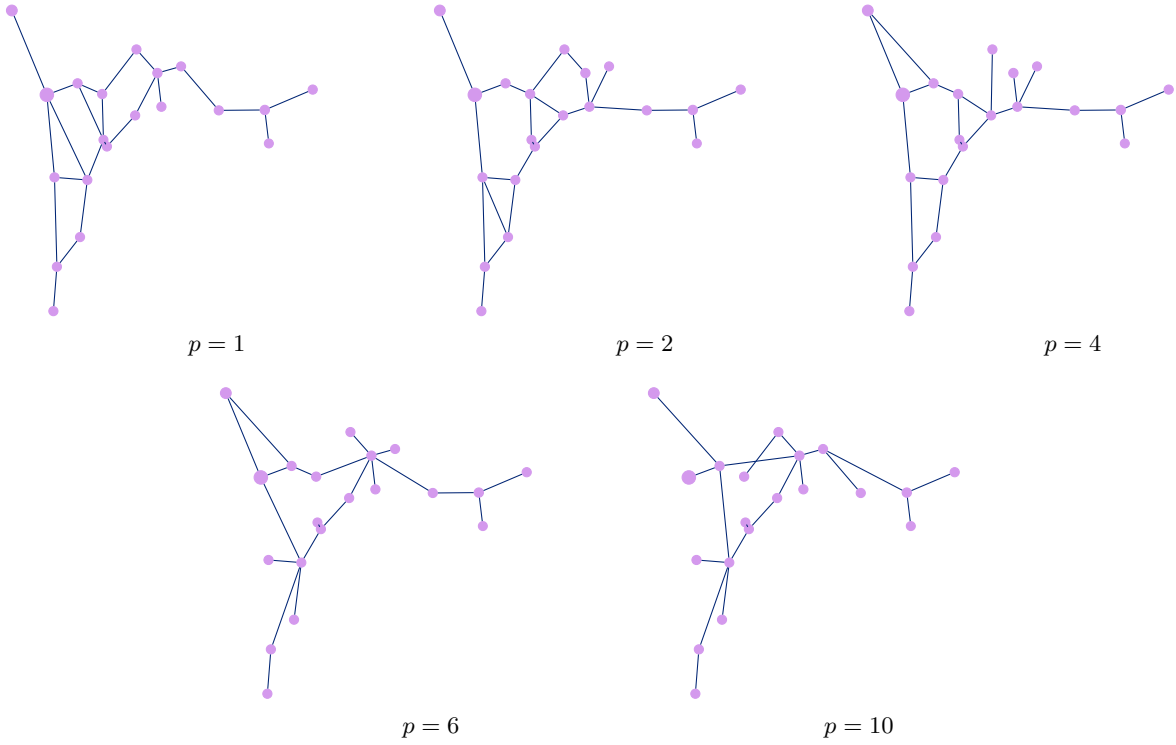

Figure 13: Solution networks for Russia with budget 9676, $K = 3$

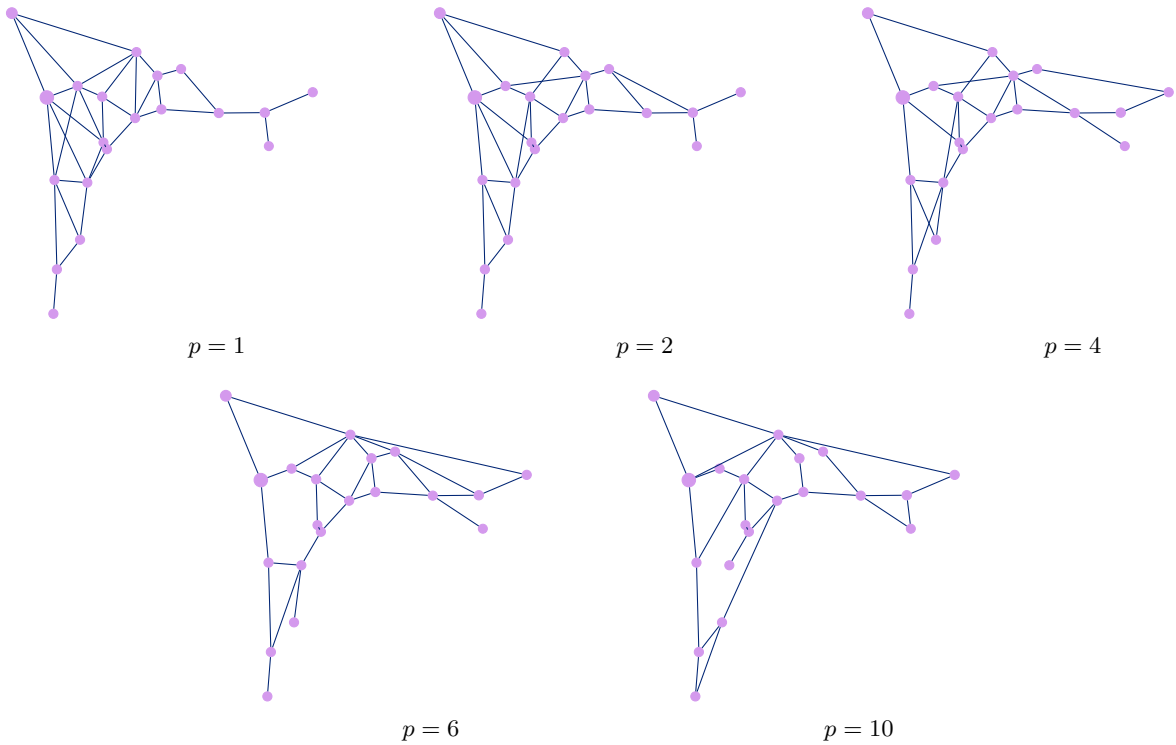

Figure 14: Solution networks for Russia with budget 17085, $K = 3$

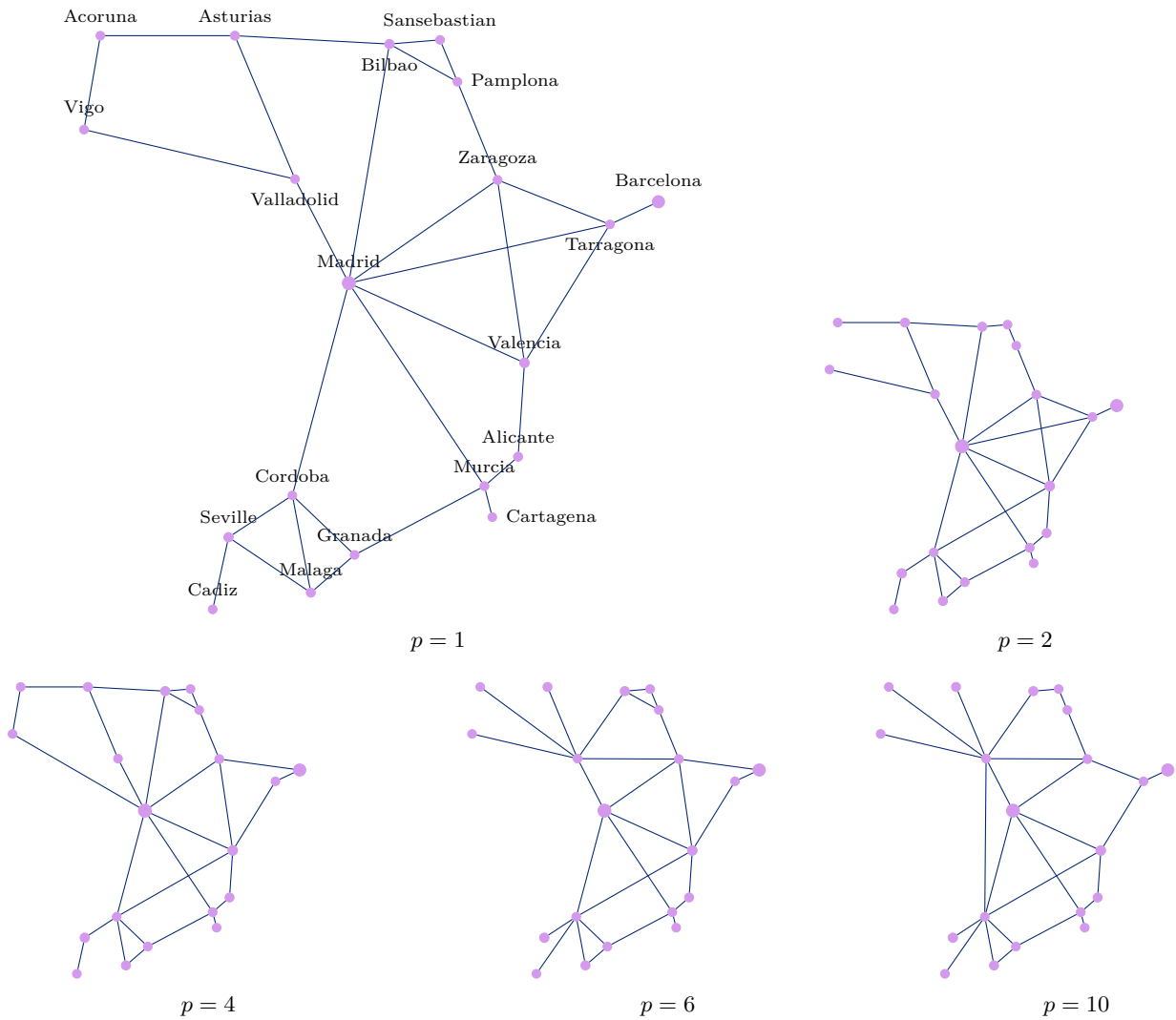

Figure 15: Solution networks for Spain with budget 5632, $K = 3$

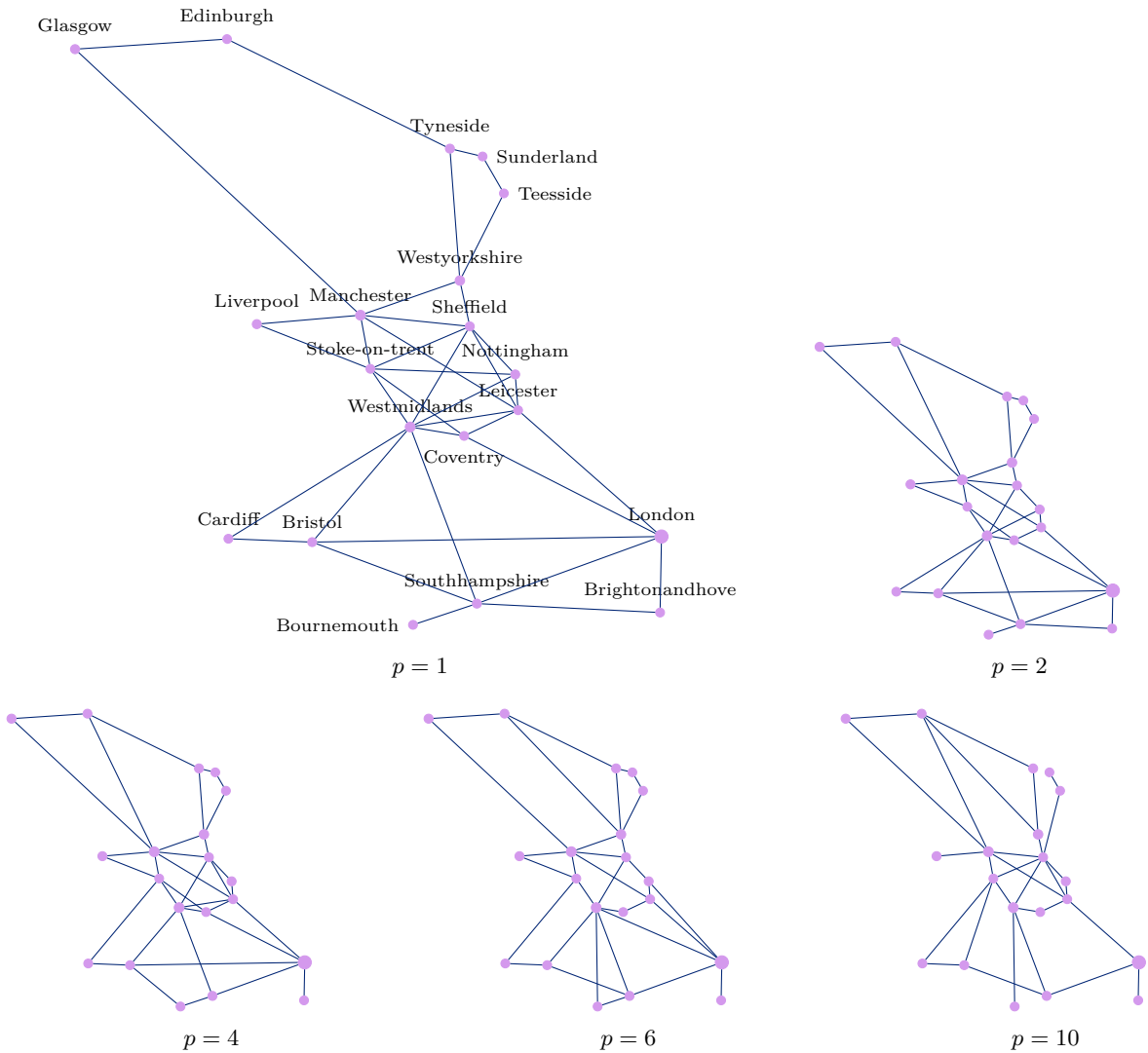

Figure 16: Solution networks for the UK with budget 3352, $K = 3$

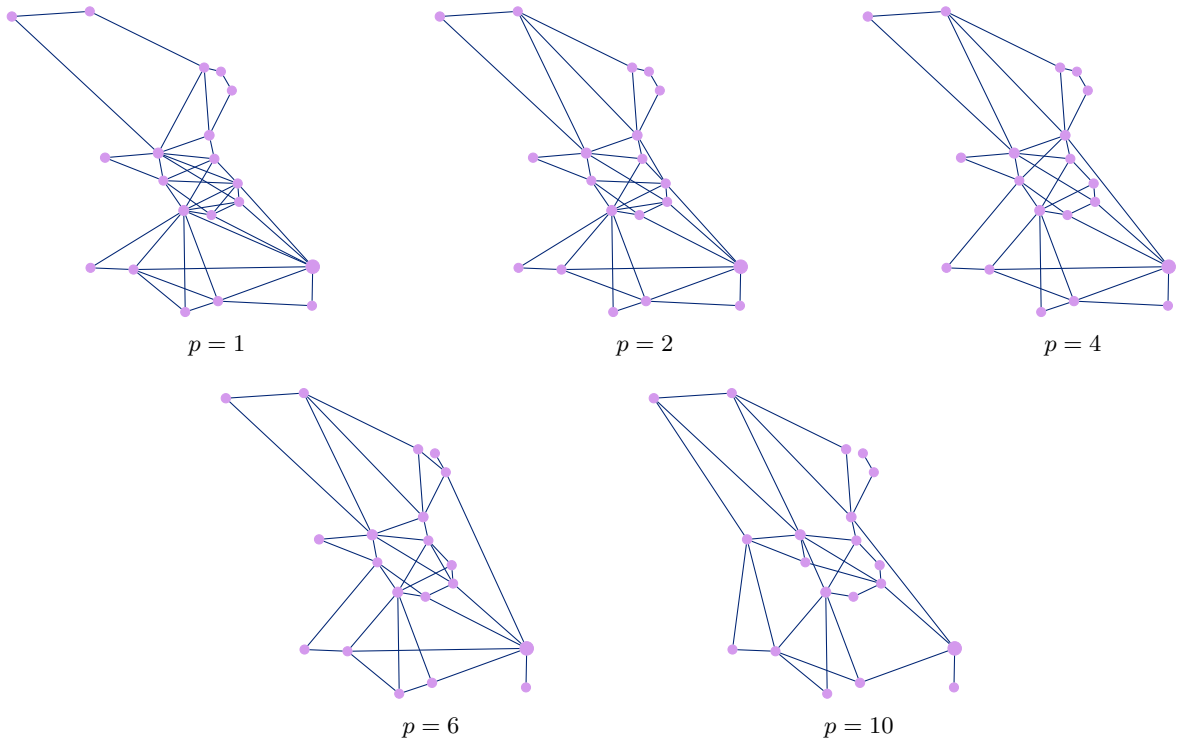

Figure 17: Solution networks for the UK with budget 4212, $K = 3$

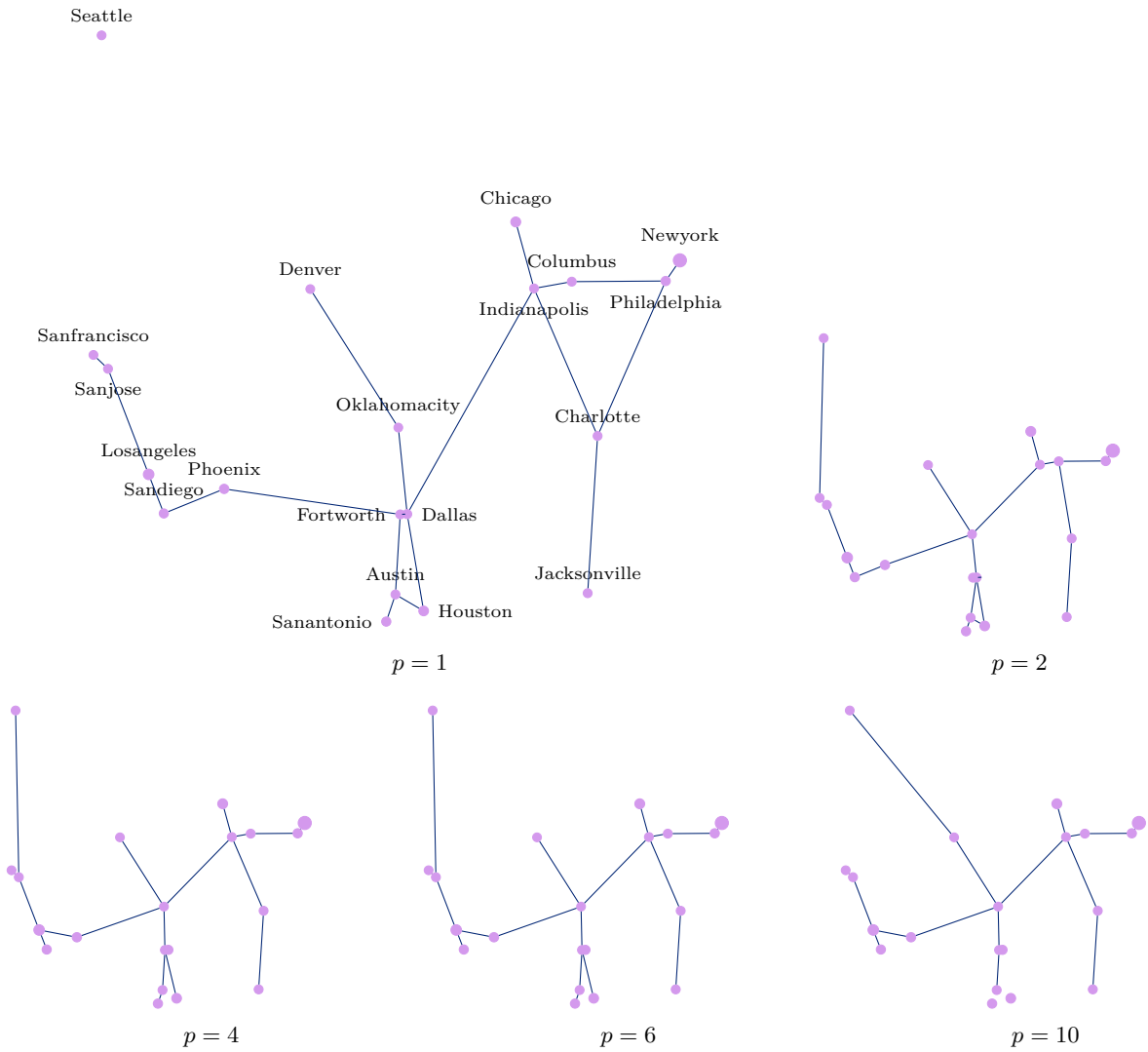

Figure 18: Solution networks for The USA with budget 9443, $K = 3$

# E  Average travel cost for all cities

Here we show, for each country considered, the average travel cost for all cities. The average is taken over all citizens traveling to/from the city and over all considered budgets. We also show the population of each city to make it easier to verify that larger cities tend to benefit more from small values of $p$ than smaller cities.

| City | Population | $p=1$ | $p=2$ | $p=4$ | $p=6$ | $p=10$ |
|---|---|---|---|---|---|---|
| Paris | 1257 | 409 | 412 | 416 | 422 | 432 |
| Lyon | 231 | 352 | 353 | 358 | 381 | 398 |
| Marseille | 176 | 414 | 409 | 412 | 425 | 440 |
| Toulouse | 135 | 582 | 570 | 568 | 586 | 607 |
| Bordeaux | 123 | 601 | 586 | 579 | 599 | 606 |
| Lille | 119 | 300 | 313 | 326 | 323 | 323 |
| Nice | 101 | 539 | 537 | 542 | 553 | 575 |
| Nantes | 96 | 477 | 480 | 483 | 494 | 512 |
| Rennes | 79 | 603 | 590 | 584 | 569 | 585 |
| Nancy | 73 | 478 | 485 | 492 | 494 | 518 |
| Grenoble | 69 | 369 | 382 | 388 | 403 | 413 |
| Rouen | 67 | 231 | 259 | 268 | 272 | 268 |
| Toulon | 63 | 354 | 354 | 357 | 367 | 385 |
| Montpellier | 61 | 421 | 416 | 426 | 452 | 467 |
| Douai | 54 | 234 | 243 | 253 | 251 | 251 |
| Avignon | 53 | 353 | 350 | 354 | 364 | 370 |
| Saint-étienne | 52 | 272 | 271 | 281 | 290 | 298 |
| Tours | 49 | 327 | 325 | 325 | 339 | 348 |
| Clermont-ferrand | 48 | 408 | 410 | 410 | 341 | 344 |

Table 1: Average travel costs: France

| City | Population | $p=1$ | $p=2$ | $p=4$ | $p=6$ | $p=10$ |
|---|---|---|---|---|---|---|
| São Paulo | 21734 | 343 | 351 | 387 | 411 | 443 |
| Rio de Janeiro | 12763 | 613 | 634 | 752 | 808 | 893 |
| Belo Horizonte | 5961 | 672 | 669 | 690 | 699 | 756 |
| Brasília | 4627 | 1046 | 1051 | 1067 | 1076 | 1101 |
| Porto Alegre | 4340 | 1350 | 1398 | 1494 | 1492 | 1463 |
| Fortaleza | 4106 | 2134 | 2068 | 2078 | 2138 | 2210 |
| Recife | 4079 | 1798 | 1742 | 1774 | 1846 | 1927 |
| Salvador | 3929 | 1508 | 1457 | 1478 | 1509 | 1605 |
| Curitiba | 3654 | 654 | 675 | 724 | 742 | 782 |
| Campinas | 3264 | 233 | 246 | 275 | 279 | 291 |
| Manaus | 2676 | 3673 | 3521 | 3272 | 3178 | 3136 |
| Goiânia | 2560 | 939 | 999 | 1044 | 1080 | 1096 |
| Vale do Paraíba | 2552 | 226 | 230 | 254 | 289 | 326 |
| Belém | 2510 | 2730 | 2612 | 2562 | 2583 | 2549 |
| Sorocaba | 2143 | 232 | 240 | 270 | 296 | 314 |
| Vitória | 1979 | 931 | 1005 | 1075 | 1113 | 1161 |
| Baixada Santista | 1865 | 174 | 184 | 211 | 220 | 227 |
| Ribeirão Preto | 1720 | 548 | 616 | 678 | 681 | 673 |
| São Luís | 1633 | 2308 | 2212 | 2172 | 2185 | 2184 |
| Natal | 1604 | 1627 | 1578 | 1633 | 1693 | 1753 |

Table 2: Average travel costs: Brazil

| City | Population | $p=1$ | $p=2$ | $p=4$ | $p=6$ | $p=10$ |
|---|---|---|---|---|---|---|
| London | 978 | 231 | 232 | 235 | 244 | 262 |
| Manchester | 255 | 160 | 162 | 169 | 167 | 180 |
| West Midlands | 244 | 153 | 157 | 170 | 163 | 180 |
| West Yorkshire | 177 | 168 | 166 | 168 | 173 | 181 |
| Glasgow | 95 | 482 | 460 | 443 | 440 | 437 |
| Liverpool | 86 | 176 | 182 | 189 | 191 | 212 |
| South Hampshire | 85 | 186 | 190 | 194 | 214 | 227 |
| Tyneside | 77 | 249 | 244 | 243 | 249 | 255 |
| Nottingham | 72 | 151 | 144 | 145 | 171 | 181 |
| Sheffield | 68 | 141 | 135 | 134 | 154 | 155 |
| Bristol | 61 | 226 | 229 | 234 | 230 | 250 |
| Leicester | 50 | 135 | 132 | 131 | 150 | 150 |
| Edinburgh | 48 | 404 | 391 | 379 | 384 | 379 |
| Brighton and Hove | 47 | 166 | 173 | 172 | 172 | 178 |
| Bournemouth | 46 | 219 | 224 | 225 | 237 | 260 |
| Cardiff | 44 | 261 | 264 | 267 | 267 | 281 |
| Teesside | 37 | 223 | 219 | 224 | 228 | 240 |
| Stoke-on-Trent | 37 | 134 | 144 | 164 | 148 | 168 |
| Coventry | 35 | 111 | 114 | 125 | 120 | 136 |
| Sunderland | 33 | 201 | 197 | 199 | 205 | 211 |

Table 3: Average travel costs: the UK

| City | Population | $p=1$ | $p=2$ | $p=4$ | $p=6$ | $p=10$ |
|---|---|---|---|---|---|---|
| Toronto | 5647 | 361 | 375 | 388 | 408 | 422 |
| Montreal | 3675 | 671 | 700 | 724 | 741 | 753 |
| Vancouver | 2426 | 1707 | 1694 | 1697 | 1729 | 1756 |
| Calgary | 1305 | 1503 | 1484 | 1486 | 1503 | 1532 |
| Edmonton | 1151 | 1574 | 1554 | 1579 | 1629 | 1687 |
| Ottawa-Gatineau | 1068 | 436 | 456 | 471 | 485 | 533 |
| Winnipeg | 758 | 1758 | 1724 | 1723 | 1723 | 1752 |
| Quebec City | 733 | 690 | 734 | 750 | 769 | 786 |
| Hamilton | 729 | 171 | 183 | 196 | 220 | 227 |
| Waterloo | 522 | 218 | 231 | 242 | 258 | 268 |
| London | 423 | 352 | 376 | 393 | 407 | 418 |
| Victoria | 363 | 711 | 712 | 727 | 779 | 799 |
| Halifax | 348 | 1638 | 1648 | 1663 | 1687 | 1706 |
| Oshawa | 335 | 150 | 158 | 162 | 165 | 175 |
| Windsor | 306 | 663 | 690 | 718 | 753 | 756 |
| Saskatoon | 264 | 1589 | 1582 | 1598 | 1639 | 1657 |
| St. Catharines-Niagara Falls | 242 | 171 | 188 | 214 | 225 | 231 |
| Regina | 224 | 1589 | 1574 | 1583 | 1589 | 1597 |
| St. John's | 185 | 2959 | 2922 | 2900 | 2877 | 2837 |
| Kelowna | 181 | 1143 | 1132 | 1132 | 1142 | 1159 |

Table 4: Average travel costs: Canada

| City | Population | $p=1$ | $p=2$ | $p=4$ | $p=6$ | $p=10$ |
|---|---|---|---|---|---|---|
| Moscow | 13010 | 1000 | 1010 | 1052 | 1109 | 1174 |
| Saint Petersburg | 5601 | 1117 | 1141 | 1195 | 1259 | 1306 |
| Novosibirsk | 1633 | 2084 | 2027 | 1974 | 1965 | 1974 |
| Yekaterinburg | 1544 | 1061 | 1072 | 1063 | 1089 | 1161 |
| Kazan | 1308 | 829 | 830 | 870 | 896 | 1025 |
| Nizhny Novgorod | 1228 | 686 | 692 | 730 | 745 | 811 |
| Chelyabinsk | 1189 | 1113 | 1092 | 1147 | 1173 | 1177 |
| Krasnoyarsk | 1187 | 2983 | 2860 | 2766 | 2756 | 2704 |
| Samara | 1173 | 794 | 806 | 846 | 870 | 909 |
| Ufa | 1144 | 1085 | 1062 | 1145 | 1166 | 1178 |
| Rostov-on-Don | 1142 | 1226 | 1249 | 1271 | 1338 | 1414 |
| Omsk | 1125 | 1795 | 1756 | 1727 | 1720 | 1739 |
| Krasnodar | 1099 | 1504 | 1529 | 1539 | 1599 | 1669 |
| Voronezh | 1057 | 836 | 858 | 888 | 973 | 1062 |
| Perm | 1034 | 1056 | 1096 | 1124 | 1226 | 1295 |
| Volgograd | 1028 | 1221 | 1245 | 1279 | 1290 | 1334 |
| Saratov | 901 | 974 | 999 | 1040 | 1043 | 1095 |
| Tyumen | 847 | 1285 | 1322 | 1276 | 1301 | 1442 |
| Tolyatti | 684 | 673 | 686 | 723 | 750 | 806 |
| Barnaul | 630 | 1844 | 1801 | 1760 | 1769 | 1788 |

Table 5: Average travel costs: Russia

| City | Population | $p=1$ | $p=2$ | $p=4$ | $p=6$ | $p=10$ |
|---|---|---|---|---|---|---|
| New York | 8258 | 1194 | 1190 | 1216 | 1229 | 1274 |
| Los Angeles | 3820 | 1532 | 1518 | 1537 | 1573 | 1630 |
| Chicago | 2664 | 1501 | 1513 | 1537 | 1566 | 1614 |
| Houston | 2314 | 1287 | 1327 | 1373 | 1448 | 1520 |
| Phoenix | 1650 | 1682 | 1660 | 1664 | 1700 | 1773 |
| Philadelphia | 1550 | 454 | 454 | 467 | 477 | 507 |
| San Antonio | 1495 | 1087 | 1120 | 1154 | 1222 | 1281 |
| San Diego | 1388 | 1031 | 1034 | 1084 | 1104 | 1157 |
| Dallas | 1302 | 797 | 810 | 815 | 846 | 882 |
| Jacksonville | 985 | 2123 | 2135 | 2200 | 2235 | 2321 |
| Austin | 979 | 865 | 891 | 917 | 976 | 1014 |
| Fort Worth | 978 | 720 | 731 | 737 | 761 | 795 |
| San Jose | 969 | 1488 | 1468 | 1456 | 1514 | 1565 |
| Columbus | 913 | 1147 | 1144 | 1170 | 1181 | 1271 |
| Charlotte | 911 | 1566 | 1578 | 1634 | 1688 | 1734 |
| Indianapolis | 879 | 1100 | 1095 | 1108 | 1120 | 1163 |
| San Francisco | 808 | 1548 | 1529 | 1518 | 1564 | 1613 |
| Seattle | 755 | 3479 | 3393 | 3343 | 3284 | 3303 |
| Denver | 716 | 2132 | 2120 | 2143 | 2117 | 2073 |
| Oklahoma City | 702 | 1375 | 1304 | 1417 | 1371 | 1389 |

Table 6: Average travel costs: the US

| City | Population | $p=1$ | $p=2$ | $p=4$ | $p=6$ | $p=10$ |
|---|---|---|---|---|---|---|
| Madrid | 615 | 446 | 449 | 454 | 460 | 457 |
| Barcelona | 517 | 519 | 517 | 523 | 531 | 543 |
| Valencia | 164 | 364 | 381 | 395 | 406 | 395 |
| Seville | 130 | 525 | 521 | 523 | 529 | 533 |
| Bilbao | 98 | 482 | 483 | 495 | 501 | 505 |
| Malaga | 94 | 500 | 506 | 507 | 513 | 518 |
| Asturias | 84 | 592 | 590 | 584 | 588 | 599 |
| Alicante | 79 | 361 | 376 | 390 | 400 | 394 |
| Zaragoza | 63 | 389 | 364 | 353 | 352 | 418 |
| Murcia | 62 | 358 | 374 | 387 | 394 | 392 |
| Granada | 44 | 466 | 469 | 468 | 467 | 474 |
| Vigo | 41 | 800 | 777 | 750 | 731 | 711 |
| Cartagena | 40 | 382 | 399 | 412 | 419 | 417 |
| Cadiz | 40 | 552 | 554 | 551 | 552 | 553 |
| San Sebastian | 39 | 433 | 427 | 438 | 450 | 463 |
| A Coruña | 37 | 760 | 751 | 729 | 725 | 727 |
| Valladolid | 36 | 362 | 361 | 361 | 359 | 353 |
| Tarragona | 32 | 244 | 243 | 248 | 255 | 259 |
| Cordoba | 31 | 400 | 397 | 396 | 396 | 396 |
| Pamplona | 28 | 416 | 402 | 403 | 408 | 446 |

Table 7: Average travel costs: Spain

| City | Population | $p=1$ | $p=2$ | $p=4$ | $p=6$ | $p=10$ |
|---|---|---|---|---|---|---|
| Berlin | 367 | 412 | 411 | 423 | 431 | 438 |
| Hamburg | 190 | 356 | 357 | 371 | 377 | 390 |
| Munich | 148 | 558 | 545 | 521 | 515 | 514 |
| Cologne | 107 | 163 | 166 | 176 | 179 | 184 |
| Frankfurt | 75 | 329 | 323 | 341 | 343 | 346 |
| Stuttgart | 62 | 460 | 452 | 450 | 454 | 453 |
| Düsseldorf | 61 | 124 | 131 | 142 | 145 | 149 |
| Leipzig | 60 | 326 | 322 | 291 | 291 | 291 |
| Dortmund | 58 | 138 | 138 | 144 | 145 | 148 |
| Essen | 57 | 107 | 110 | 116 | 118 | 120 |
| Bremen | 56 | 306 | 311 | 318 | 319 | 339 |
| Dresden | 55 | 395 | 390 | 382 | 394 | 397 |
| Hanover | 53 | 256 | 253 | 255 | 258 | 259 |
| Nuremberg | 51 | 390 | 379 | 360 | 359 | 359 |
| Duisburg | 49 | 117 | 124 | 134 | 135 | 137 |
| Bochum | 36 | 102 | 104 | 109 | 111 | 113 |
| Wuppertal | 35 | 117 | 118 | 124 | 125 | 127 |
| Bielefeld | 33 | 234 | 233 | 237 | 240 | 242 |
| Bonn | 33 | 156 | 159 | 169 | 172 | 177 |
| Münster | 31 | 214 | 224 | 234 | 231 | 233 |

Table 8: Average travel costs: Germany

| City | Population | $p=1$ | $p=2$ | $p=4$ | $p=6$ | $p=10$ |
|---|---|---|---|---|---|---|
| Rome | 422 | 400 | 395 | 395 | 397 | 407 |
| Milan | 322 | 152 | 155 | 162 | 171 | 184 |
| Naples | 298 | 285 | 280 | 280 | 280 | 290 |
| Turin | 220 | 388 | 392 | 401 | 416 | 431 |
| Brescia | 125 | 218 | 224 | 246 | 262 | 286 |
| Bari | 122 | 540 | 540 | 541 | 544 | 555 |
| Palermo | 120 | 849 | 818 | 804 | 794 | 787 |
| Bergamo | 110 | 186 | 192 | 194 | 199 | 212 |
| Catania | 107 | 508 | 495 | 486 | 482 | 479 |
| Salerno | 106 | 253 | 247 | 248 | 248 | 259 |
| Bologna | 101 | 275 | 282 | 293 | 303 | 317 |
| Florence | 98 | 325 | 324 | 325 | 328 | 338 |
| Padua | 93 | 192 | 196 | 210 | 220 | 241 |
| Verona | 92 | 99 | 102 | 110 | 116 | 129 |
| Caserta | 90 | 217 | 217 | 222 | 226 | 231 |
| Varese | 87 | 169 | 174 | 185 | 195 | 215 |
| Treviso | 87 | 293 | 302 | 320 | 331 | 358 |
| Monza e Brianza | 87 | 84 | 87 | 92 | 98 | 105 |
| Vicenza | 85 | 95 | 97 | 104 | 110 | 122 |
| Messina | 60 | 407 | 395 | 388 | 384 | 382 |

Table 9: Average travel costs: Italy